\documentclass{PoS}

\title{Lattice study of nuclear forces}

\ShortTitle{Lattice study of nuclear forces}

\author{\speaker{Noriyoshi Ishii}\\
        Department of Physics, The university of Tokyo\\
        E-mail: \email{ishii@ribf.riken.jp}}

\author{for PACS-CS and HAL QCD Collaborations}
\author{\includegraphics[width=0.20\textwidth]{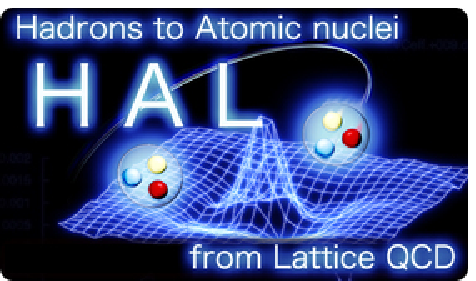}}

\abstract{  Recent progress  of lattice  QCD study  of  nuclear forces
  (potentials) is reviewed.
Scattering  phase shift is  an important  observable for  two particle
system.   In  lattice  QCD,  phase  shifts are  calculated  from  long
distance behavior of Bethe-Salpeter (BS) wave functions by L\"uscher's
finite  volume  method.   For   applications  to  nuclear  physics  of
multi-nucleon  system,   it  is  more  advantageous   to  convert  the
information of phase  shifts in the form of  potentials.  We therefore
extend  the method  so  as to  generate  the potentials  from BS  wave
functions.  These potentials are faithful to scattering phase shift by
construction, because  they can reproduce  BS wave functions  in which
the  information  of phase  shift  is  embeded  in the  long  distance
part.The  method was  first applied  to  the central  potential in  NN
system.  It is now applied  to many objects, such as tensor potential,
hyperon  potentials,  energy  dependence  of nuclear  potentials,  and
investigations of the repulsive core at short distance.
}

\FullConference{The XXVII International Symposium on Lattice Field Theory - LAT2009\\
		 July 26-31 2009\\
		 Peking University, Beijing, China}

\newcommand{\Eq}[1]{Eq.~(\ref{#1})}
\newcommand{\Fig}[1]{Fig.~\ref{#1}}
\newcommand{\Sect}[1]{Sect.~\ref{#1}}
\newcommand{\Ref}[1]{Ref.~\cite{#1}}
\newcommand{\Hs}{\hspace*{1em}}
\newcommand{\Tate}{\rule{0cm}{1.8ex}}
 
\newcommand{\agt}{\;\raisebox{.5ex}{$>$}\hspace{-0.8em}\raisebox{-.7ex}{$\sim$}\;}
\newcommand{\alt}{\;\raisebox{.5ex}{$<$}\hspace{-0.8em}\raisebox{-.7ex}{$\sim$}\;}

\begin{document}

\section{Introduction}

Scattering phase shift is an  important physical observable in its own
right.
%%%...................................................................
However, for  nuclear physics, it  is more advantageous to  convert it
into a form of the nuclear potential.
%%%...................................................................
Once such a nuclear potential  is at our disposal, we can conveniently
use it to study a variety  of nuclear phenomena based on the effective
degrees of freedom,  i.e., the nucleons.  It provides  us with physics
insights into  the structures and  the reactions of atomic  nuclei, as
well  as the  supernova  explosion of  type  II and  the structure  of
neutron stars through the equation of states of cold and dense nuclear
matter.
%%%...................................................................
Enormous efforts in this line  are integrated into a form of realistic
nuclear  potentials  \cite{realistic.nuclear.force}.   By using  about
$40-50$ adjustable parameters, they can reproduce several thousands of
experimental NN data with $\chi^2/\mbox{NDF} \sim 1$, which consist of
the scattering phase shifts and the deuteron property.
%%%...................................................................
%% There  exist   several  versions  of   realistic  nuclear  potentials.
%% However, it is  important to keep in mind that,  even if their precise
%% shapes  are different  from  each  other, they  gives  the same  phase
%% shifts.
%% %%%...................................................................
Also, the potentials from the chiral effective field theory attract an
growing interest \cite{chiral.eft}.

Unlike  the nucleon  sector,  only a  limited  number of  experimental
information is  available in the hyperon  sector.  This is  due to the
absence of  accelerator facilities, which can  generate direct hyperon
beam.
%%%...................................................................
If there were  a sufficient number of experimental  information on the
hyperon-hyperon  and   hyperon-nucleon  scatterings,  they   would  be
converted into realistic hyperon potentials, which could help us study
the  structure  of hyper-nuclei  and  possible  generation of  hyperon
matter in the neutron star core.

The standard method  to obtain the scattering phase  shifts in lattice
QCD is L\"uscher's finite volume method \cite{luescher}.
%%%...................................................................
It  can  be  used  to  provide QCD  predictions/postdictions  for  the
scattering   phase   shifts   not   only   in   the   nucleon   sector
\cite{fukugita,beane}, but also in the hyperon sectors \cite{nplqcd}.
%%%...................................................................
One  may  come up  with  a  straightforward  way to  obtain  realistic
inter-baryon  potentials in  lattice QCD,  i.e., sufficient  number of
scattering  phase shifts are  generated by  L\"uscher's method  at the
initial stage, which are converted to inter-baryon potentials with the
help of the inverse scattering theory.
%%%...................................................................
However,  this  is  difficult  in  practice, because  it  involves  an
infinite number of scattering phase shift at the initial stage.
%%%...................................................................
Thus,  it is desirable  to have  a direct  method to  obtain realistic
inter-baryon potentials in lattice QCD.
%%%...................................................................

The method recently  proposed by \Ref{ishii} is such  a method. It can
be given a background in terms of L\"uscher's finite volume method.
%%%...................................................................
By using  the effective Schr\"odinger equation,  it constructs nuclear
potentials from  the Bethe-Salpeter  (BS) wave functions  generated by
lattice QCD.
%%%...................................................................
Since the information of the scattering phase shift is embedded in the
long  distance  part of  the  BS wave  functions,  it  is possible  to
generate  a  realistic nuclear  potential,  which  reproduces the  QCD
predictions  of the  scattering phase  shift extracted  by L\"uscher's
method.

In  this paper,  after  a brief  review  of the  general  idea how  to
construct  a realistic nuclear  potential in  lattice QCD,  we present
quenched QCD results of the central and the tensor potentials obtained
at the leading order of the derivative expansion.
%%%...................................................................
After the  discussion of the convergence of  the derivative expansion,
we give dynamical QCD results by using 2+1 flavor gauge configurations
generated by PACS-CS Collaboration.
%%%...................................................................
Finally, we mention the hyperon potentials ($N\Xi$ and $N\Lambda$), to
which our method can be equally applied.

\section{General idea to construct nuclear potential in QCD}

We consider  (equial-time) Bethe-Salpeter  (BS) wave function  for two
nucleons  in  the  center  of  mass  frame.   By  choosing  particular
composite  interpolating  fields  $p(x)$  and $n(x)$  for  proton  and
neutron, the BS wave function is defined as
\begin{equation}
  \psi_{\vec k}(\vec x - \vec y)
  \equiv
  \left\langle 0 \left| \Tate
  p(\vec x) n(\vec y)
  \right| p(\vec k)n(-\vec k),\mbox{in}\right\rangle
%  \left(
  =
  \lim_{t\to +0}
  \left\langle 0 \left| \Tate
  T\left[ p(\vec x,t) n(\vec y,0) \right]
  \right| p(\vec k)n(-\vec k),\mbox{in}\right\rangle,
%  \right),
\end{equation}
where $\vec k$ denotes the  asymptotic momentum of the proton relative
to the neutron.  It is related to the relativistic total energy of the
state as $P_0 = 2\sqrt{m_N^2 + \vec k^2}$ with $m_N$ being the nucleon
mass.
%%%%%%%%%%%%%%%%%%%%%%%%%%%%%%%%%%%%%%%%%%%%%%%%%%%%%%%%%%%%%%%%%%%%%%
Quite naively, this matrix element  may be regarded as an amplitude to
find three  quarks at $\vec x$  and another three quarks  at $\vec y$,
where $p(\vec  x)$ and $n(\vec y)$  are used to probe  nucleons in the
state $|p(\vec k) n(-\vec k),\mbox{in}\rangle$.
%%%%%%%%%%%%%%%%%%%%%%%%%%%%%%%%%%%%%%%%%%%%%%%%%%%%%%%%%%%%%%%%%%%%%%
At  long distance $|\vec  x -  \vec y|\to  \mbox{large}$, it  shows an
asymptotic behavior,  which is  characterized by the  scattering phase
shift  $\delta_l(k)$ in  exactly the  same  way as  a scattering  wave
function in the quantum mechanics as \cite{lin,ishizuka,hal}
\begin{equation}
  \psi_{\vec k}(\vec x - \vec y)
  \simeq
  A
  \frac{
    \sin(k|\vec x - \vec y| - \pi l / 2 + \delta_l(k))
  }{
    k|\vec x - \vec y|
  }
  +\cdots.
  \label{asymptotic.form}
\end{equation}
%%%...................................................................
To  prove  this  behavior,  Nishijima-Zimmerman-Haag  (NZH)  reduction
formula  \cite{nishijima}  is conveniently used.
%%%...................................................................
Note that any  local composite nucleon field $N(x)$  leads to the same
asymptotic  behavior  \Eq{asymptotic.form},   as  far  as  $N(x)$  has
non-vanishing overlap  with a single nucleon state,  i.e., $\langle 0|
N(x) | N\rangle \neq 0$.
%%%...................................................................
In L\"uscher's finite  volume method, which is the  standard method to
calculate the scattering phase shift in lattice QCD,
%%%...................................................................
the  phase  shift embedded  in  BS wave  function  in  this manner  is
extracted  from   the  energy  spectrum  in  a   finite  periodic  box
\cite{luescher}.  (For explicit use of BS wave function in L\"uscher's
finite volume method, see \Ref{ishizuka}.)

For nuclear  physics, it is more  advantageous to convert  the data of
the phase shift  into a form of nuclear  potentials. We therefore wish
to extend  L\"uscher's method so  as to obtain the  nuclear potentials
directly.
%%%...................................................................
For this purpose,  we use the remarkable similarity  in the asymptotic
behaviors between the BS wave function \Eq{asymptotic.form} of QCD and
the scattering wave function in the quantum mechanics.
%%%...................................................................
%% The asymptotic behavior \Eq{asymptotic.form} of BS wave function is
%% analogous  to  the  non-relativistic  wave function  of  scattering
%% states in the quantum mechanics.
%%%...................................................................
This similarity motivates us to  construct a nuclear potential so that
it  can reproduce  all the  BS wave  functions simultaneously  in wide
range of energy region.
%%%%%%%%%%%%%%%%%%%%%%%%%%%%%%%%%%%%%%%%%%%%%%%%%%%%%%%%%%%%%%%%%%%%%%
Then, the resulting potential can reproduce the phase shifts predicted
by  QCD. In  this  way,  it becomes  possible  to construct  realistic
nuclear potentials by lattice QCD.

To proceed, we define the nuclear potential $U(\vec r,\vec r')$ by the
effective Schr\"odinger equation \cite{hal}
\begin{equation}
  (\triangle + k^2)
  \psi_{\vec k}(\vec r)
  =
  m_N
  \int d^3 r'\;
  U(\vec r,\vec r')
  \psi_{\vec k}(\vec r'),
  \label{eq.effective.schrodinger}
\end{equation}
where $m_N$ denotes the nucleon mass.
%% where $\mu\equiv  m_N/2$ denotes the  reduced mass of the  two nucleon
%% system with the nucleon mass $m_N$.
%%%...................................................................
We demand this equation  to be simultaneously satisfied by $\psi_{\vec
  k}(\vec r)$ in wide $\vec  k$ region (or wide energy $2\sqrt{m_N^2 +
  \vec k^2}$ region).
%%%...................................................................
Note that $U(\vec r,\vec r')$ is most generally a non-local potential,
and that, with our definition,  $U(\vec r,\vec r')$ does not depend on
the relativistic total energy $P_0\equiv 2\sqrt{m_N^2 + \vec k^2}$.

Several comments are in order.

(i)  With our  prescription, precise  forms of  potentials  depend on
particular choices of interpolating fields $p(x)$ and $n(x)$. However,
even if  their particular shapes are different,  these potentials lead
to  the  same  phase   shift.   Remember  that  these  potentials  are
constructed  so as  to give  the phase  shift obtained  by L\"uscher's
method, which does not depend  on a particular choice of interpolating
fields.
%%%...................................................................
The situation  is analogous to  the unitary transformation  in quantum
mechanics,  i.e.,   shapes  of  potentials  are   changed  by  unitary
transformations without affecting any observables.
%%%...................................................................
The potentials,  which lead to  the same phase  shift, are said  to be
phase-shift equivalent to each other.

%% (ii)  In  general,  the BS  wave  functions  are  not expected  to  be
%% orthogonal  to  each  other  with  respect  to  the  three-dimensional
%% integral    as    $\displaystyle    \int    d^3    x\;\;    \psi_{\vec
%% k';\alpha\beta}^*(\vec x) \psi_{\vec k;\alpha\beta}(\vec x) \neq {\cal
%% N} \delta^3(\vec k - \vec  k')$.
%% %%%...................................................................
%% If the  violation of orthogonality is severe,
%% %%%...................................................................
%% we should  orthogonalize BS wave functions before  using the procedure
%% above in  order to avoid  the non-Harmitian nuclear  potentials.  (See
%% \Ref{hal} for detail.)
%%%...................................................................
(ii)  Here,   for  simplicity,  we   do  not  pay  attention   to  the
orthogonality  of the  BS wave  functions.  In  general, the  BS wave
functions are not expected to be orthogonal to each other with respect
to  the three-dimensional  integral as  $\displaystyle \int  d^3 x\;\;
\psi_{\vec k';\alpha\beta}^*(\vec x) \psi_{\vec k;\alpha\beta}(\vec x)
\neq {\cal  N} \delta^3(\vec k  - \vec k')$.  If the violation  of the
orthogonality   is   serious,   we   have   to  take   care   of   the
orthogonalization to avoid non-hermitian potentials \cite{hal}.

\section{The Derivative Expansion}

To  construct   the  non-local   potential  $U(\vec  r,\vec   r')$  in
\Eq{eq.effective.schrodinger},  it is  necessary to  generate infinite
number of BS wave functions, which is difficult in lattice QCD.
%%%...................................................................
The reason is  two fold.  (i) The energy spectrum  is discretized in a
finite spatial box.  (ii) Although there is a method to access excited
states in  lattice QCD,  it becomes the  more difficult to  access the
higher excited states.
%%%...................................................................
Thus, we need an approximation,  which enables us to construct $U(\vec
x,\vec x')$ with a limited number of BS wave functions.
%
%% For this purpose,  we use the derivative expansion,  by which possible
%% non-locality can  be taken into account  step by step  starting from a
%% local potential.
%%%...................................................................
For this purpose, we use  the derivative expansion.  We can start with
the  leading local  potentials, and  then take  into  account possible
non-local  terms  (potentials,  which  contain derivatives)  order  by
order.   If   the  non-locality  effect  appears  to   be  large,  the
convergence  can  be improved  by  changing  the interpolating  fields
$p(x)$ and $n(y)$ in the sink side.

To proceed, we impose  general requirements on the non-local potential
$U(\vec x,\vec x')$
arising  from  the   translational  invariance,  Galilean  invariance,
symmetry condition  (identical particle condition),  spatial rotation,
spatial reflection, time-reversal invariance and hermiticity.
The most  general form  has been derived  in \Ref{okubo}, to  which we
apply the derivative expansion. We are left with
\begin{eqnarray}
  U^{I}(\vec x,\vec x')
  &=&
  V_{NN}^{I}(\vec x,\vec \nabla) \delta^{3}(\vec x - \vec x')
  \\\nonumber
  V_{NN}^{I}(\vec x,\vec \nabla)
  &=&
  V_{0}^{I}(r)
  + V_{\sigma}^{I}(r)\;
  \vec \sigma_1\cdot \vec \sigma_2
  + V_{T}^{I}(r)\;
  S_{12}
  + V_{LS}^{I}(r)\;
  \vec L\cdot \vec S
  + O(\nabla^2),
\end{eqnarray}
where  $I$ indicates  the total  iso-spin of  the two  nucleon system.
$\sigma_1$ and $\sigma_2$ act on the spin indices of the first and the
second nucleons,  respectively.  $S_{12} \equiv  3(\vec \sigma_1 \cdot
\vec x) (\vec \sigma_2 \cdot \vec  x) / \vec x^2 - \vec \sigma_1 \cdot
\vec \sigma_2$ is referred to  as the tensor operator.  $\vec L \equiv
i\; \vec  x \times \vec  \nabla$ denotes the orbital  angular momentum
operator. $\vec  S \equiv (\vec  \sigma_1 + \vec  \sigma_2)/2$ denotes
the total spin operator.
%%%...................................................................
Since  $\vec \sigma_1  \cdot \vec  \sigma_2$ reduces  to $1$  for spin
triplet, and  $-3$ for spin  singlet, $V_0(r) + V_{\sigma}(r)  \; \vec
\sigma_1 \cdot  \vec \sigma_2$ is conveniently combined  into the form
of the ``{\em central potential}'' $V_{C}(r)$ as
\begin{equation}
  V_{C}(r; ^1S_0)
  \equiv
  V_0(r) - 3 V_{\sigma}(r),
  \Hs\Hs\Hs
  V_{C}(r; ^3S_1)
  \equiv
  V_0(r) + V_{\sigma}(r).
\end{equation}
$V_{T}(r)$ and $V_{LS}(r)$ are referred to as the tensor potential and
the LS  potential, respectively. Note that  $V_{C}(r)$, $V_{T}(r)$ and
$V_{LS}(r)$ play important roles in conventional nuclear physics.

At the leading order, we  truncate the nuclear potential by neglecting
all the derivative terms as
\begin{equation}
  V_{NN}(\vec x,\vec \nabla)
  =
  V_{C}(r)
  +
  V_{T}(r)\;
  S_{12}
  +
  O(\vec \nabla).
  \label{eq.leading.potential}
\end{equation}
We   insert   this   into   the   effective   Schr\"odinger   equation
\Eq{eq.effective.schrodinger} to have
\begin{equation}
  \left(
  -\frac{\triangle}{m_N} + V_{C}(r) + V_{T}(r)\;S_{12}
  \right)
  \psi_{\vec k}(\vec r)
  =
  \frac{k^2}{m_N}
  \psi_{\vec k}(\vec r).
  \label{eq.leading.schrodinger}
\end{equation}
Note  that the  reduced  mass of  two  nucleon system  is $\mu  \equiv
m_N/2$. For $J^{P}=0^+$ ($^1S_0$),  $I=1$ channel, a further reduction
is possible.  Since $S_{12}$ acts as zero on wave functions in $^1S_0$
channel, we can rewrite it as
\begin{equation}
  V_{C}(r; ^1S_0)
  =
  \frac{k^2}{m_N}
  +
  \frac1{m_N}
  \frac{\triangle \psi_{\vec k}(\vec r)}{\psi_{\vec k}(\vec r)}.
  \label{eq.leading.central.1s0}
\end{equation}
We use this formula to obtain the central potential in $^1S_0$ channel
at the leading order of the derivative expansion.
For  $J^{P}=1^+$  ($^3S_1-^3D_1$ coupled  system)  $I=0$ channel  (the
deuteron channel), the procedure is  slightly involved, which is to be
considered in \Sect{section.tensor.potential}.
%
%
%
%% Note that, to obtain these strictly local potentials at leading order,
%% we  need only to  use a  BS wave  function for  a single  energy eigen
%% state.
  
At the next  to leading order, we include the  terms, which contains a
single derivative, i.e., $V_{LS}(r)\; \vec L \cdot \vec S$ as
\begin{equation}
  V_{NN}(\vec x,\vec \nabla)
  =
  V_{C}(r)
  +
  V_{T}(r)\;
  S_{12}
  +
  V_{LS}(r)\;
  \vec L \cdot \vec S
  +
  O(\vec \nabla^2).
\end{equation}
This   is   inserted  into   the   effective  Schr\"odinger   equation
\Eq{eq.effective.schrodinger}. Note  that the action of  $\vec L \cdot
\vec  S$  on $J^P=0^+$  ($^1S_0$)  channel  vanishes.  Therefore,  the
formula \Eq{eq.leading.central.1s0} does not change at this order.  In
contrast, $\vec  L\cdot \vec S$ gives a  non-vanishing contribution to
$J^P=1^+$  ($^3S_1-^3D_1$) channel.  Hence,  the formula  to calculate
$V_{C}(r)$ and $V_{T}(r)$ is modified.
We  need  an  additional  BS  wave  function  to  obtain  these  three
potentials at this order.

At the  next to next to  leading order, we  include $O(\vec \nabla^2)$
terms  in   the  potential,  which   is  inserted  to   the  effective
Schr\"odinger equation \Eq{eq.effective.schrodinger}.
To obtain these potentials, we need further BS wave functions.
We perform this procedure repeatedly to obtain higher derivative terms
by using increasing number of BS wave functions.

It  is  important  to   examine  the  convergence  of  the  derivative
expansion.
The non-local  potential is  faithful to the  scattering data  in wide
range of  energy region, while it  may not be so  after the derivative
expansion is applied.
%%%...................................................................
If the convergence appears to be unsatisfactory, improvement has to be
done by changing interpolating field of nucleon.

\section{Lattice QCD setup}

We use quenched QCD unless otherwise indicated.
We  employ the  standard plaquette  gauge action  with  $\beta=5.7$ to
generate gauge  configurations on the lattice of  the size $32^3\times
N_t$ ($N_t =  32$ and $48$).
The scale  unit is introduced by  rho meson mass in  the chiral limit,
which  leads to  the  lattice spacing  $a^{-1}=1.44(2)$ GeV  ($a\simeq
0.137$ fm) \cite{fukugita}.  The spatial extension amounts to $L = 32a
\simeq 4.4$ fm.
1000-4000 gauge configurations are used in our calculations.
Quark propagators are generated by employing the standard Wilson quark
action  with  the   hopping  parameters  $\kappa=0.1640,  0.1665$  and
$0.1678$, which correspond to $m_{\pi}  \simeq 731, 529$ and $380$ MeV
and $m_{N} \simeq 1558, 1334$ and $1197$ MeV, respectively.
Unless otherwise indicated, Dirichlet and periodic boundary conditions
are imposed on quark fields along the temporal and spatial directions,
respectively.

To obtain the BS wave  function, we generate the four point correlator
of the nucleon field as
\begin{equation}
  G(\vec x - \vec y, t)
  \equiv
  \left\langle 0 \left|
  T\left[\Tate
    p(\vec x,t)
    n(\vec y,t)
    \overline{\bf p}(t=0)
    \overline{\bf n}(t=0)
    \right]
  \right| 0 \right\rangle,
  \label{eq.four.point.function}
\end{equation}
where $p(x)$ and  $n(x)$ denote local composite fields  for proton and
neutron, for which we employ the standard ones as
%%%...................................................................
$p(x)
  \equiv
  \epsilon_{abc} \left(  u^T_a(x) C\gamma_5 d_b(x) \right) u(x),
$
%%%...................................................................
$n(x)
  \equiv
  \epsilon_{abc}  \left( u^T_a(x) C\gamma_5 d_b(x) \right) d(x).$
%%%...................................................................
These fields are represented  by Heisenberg picture in imaginary time.
$\overline{\bf p}(t)$  and $\overline{\bf n}(t)$  denote interpolating
fields for proton and neutron (wall source), i.e.,
%%%...................................................................
$\overline{\bf p}(t)
  \equiv
  \sum_{\vec x,\vec y,\vec z}
  \epsilon_{abc}
  \overline{u}_c(\vec x,t)
  \left( \overline{d}_b(\vec y,t) C\gamma_5 \overline{u}_a(\vec z,t)^T \right),
$
%%%...................................................................
$
  \overline{\bf n}(t)
  \equiv
  \sum_{\vec x,\vec y,\vec z}
  \epsilon_{abc}
  \overline{d}_c(\vec x,t)
  \left( \overline{d}_b(\vec y,t) C\gamma_5 \overline{u}_a(\vec z,t)^T \right).
$
%%%...................................................................
Note that the  total spatial momentum of the  system vanishes because
of the  wall source.  Therefore  \Eq{eq.four.point.function} becomes a
function of $\vec x - \vec y$ due to the translational invariance.
\Eq{eq.four.point.function} for  the large Euclidean  time $t >  0$ is
dominated by the contribution from the lowest-lying state as
\begin{equation}
  G(\vec x - \vec y, t)
  =
  \sum_{n}
  \left\langle 0 \left|\Tate
  p(\vec x,t)
  n(\vec y,t)
  \right| E_n \right\rangle
  \left\langle E_n \left|\Tate
  \overline{\bf p}(0)
  \overline{\bf n}(0)
  \right| 0 \right\rangle
  =
  A_0\;
  e^{-E_0 t}
  \psi_{E_0}(\vec x - \vec y)
  +
  \cdots,
\end{equation}
where  $E_0$  denotes  the  energy   of  the  lowest  lying  state  $|
E_0\rangle$,  and   $A_0  \equiv  \langle   E_0|  \overline{\bf  p}(0)
\overline{\bf n}(0)| 0 \rangle$.
%%%...................................................................
$\psi_{E_0}(\vec x - \vec  y) \equiv \left\langle 0 \left|\Tate p(\vec
x) n(\vec y) \right| E_0  \right\rangle$ denotes the BS wave function.
Needless to say, $\psi_{E_0}(\vec r)$  does not depend on a particular
choice  of the interpolating  fields in  the source  side, as  long as
$\langle  E_0| \overline{\bf  p}\;  \overline{\bf n}  | 0\rangle  \neq
0$. (We use the wall source for the efficiency reason.)
Quantum numbers of the wave  function such as $J^{P}$ are controlled by
quantum  numbers  of the  interpolating  fields  in  the source  side.
Because  we adopted  the  wall source,  $J^{P}  = 0^+$  and $1^+$  are
obtained   by  combining   the   spins  of   $\overline{\bf  p}$   and
$\overline{\bf n}$.

\section{Central potential at the leading order of the derivative expansion}

\begin{figure}[h]
\begin{center}
  \includegraphics[height=0.48\textwidth,angle=-90]{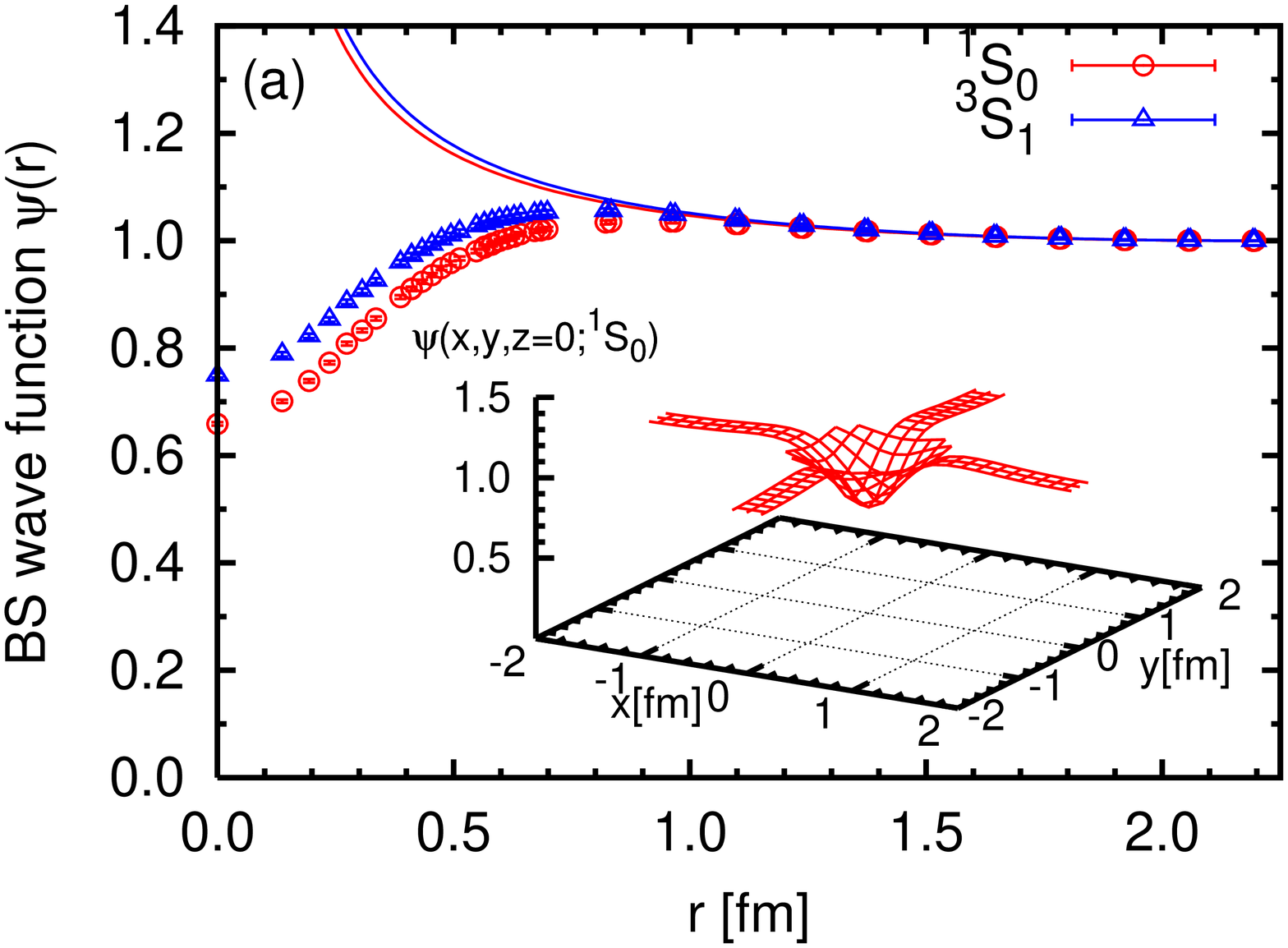}
  \includegraphics[height=0.48\textwidth,angle=-90]{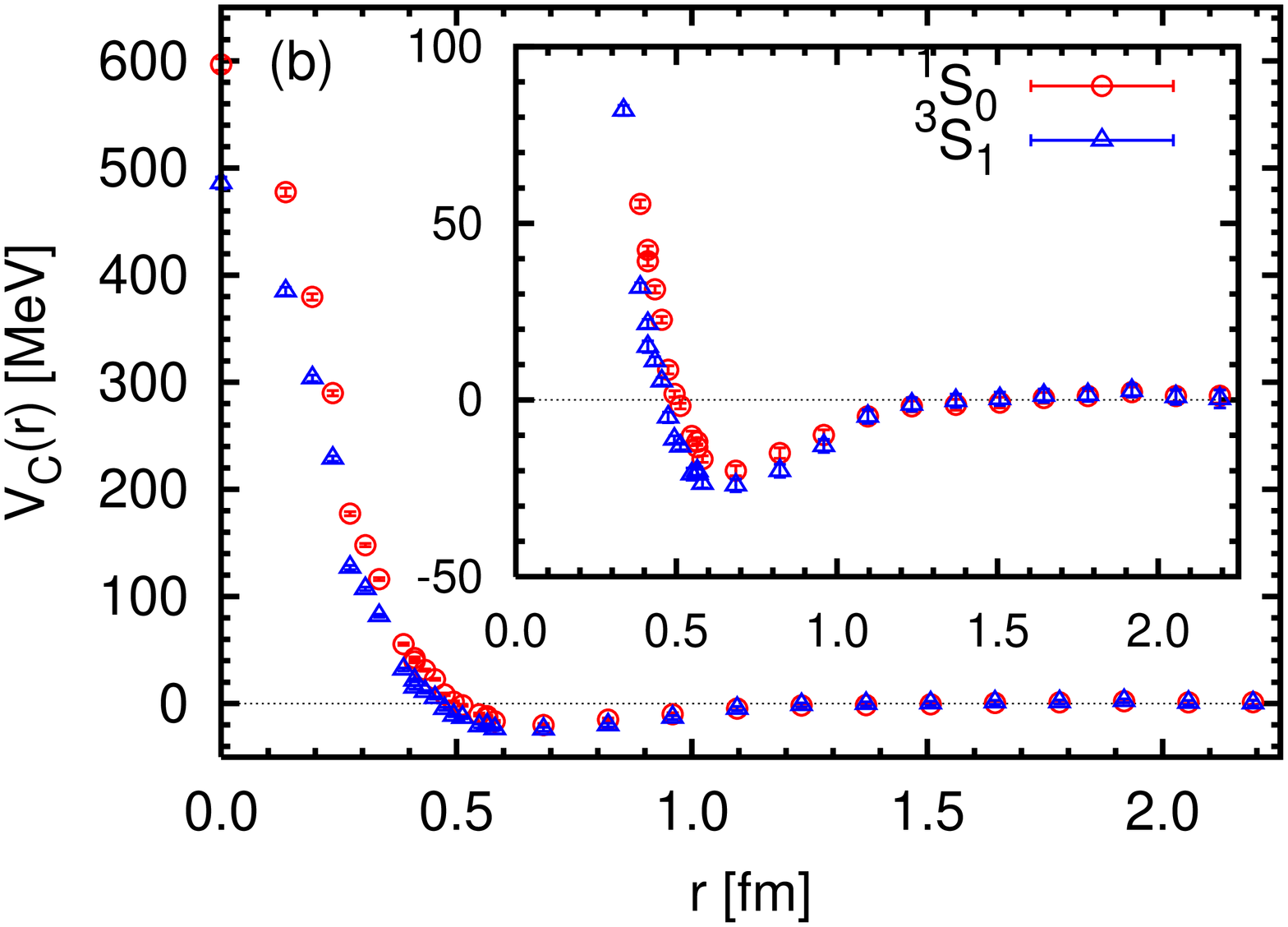}
\end{center}
\caption{(a)  BS wave functions  in $^1S_0$  and $^3S_1$  channels for
  $m_{\pi}\simeq 529$ MeV.  The inset is a 3D  plot of $\psi(x,y,z=0)$
  for $^1S_0$.  The curves  denote the results  of the fits  using the
  Green's function of Helmholtz equation  in the region $11a \le r \le
  16a$  and (b)  The  central  potential in  $^1S_0$  channel and  the
  effective  central potential in  $^3S_1$ channel  for $m_{\pi}\simeq
  529$ MeV.}
\label{fig.s-wave}
\end{figure}
\Fig{fig.s-wave}(a) shows the quenched  result of BS wave functions in
$^1S_0$ and $^3S_1$ channels.
%%%...................................................................
To  pick up  ``{\em s-wave}''  component  ($^3S_1$) from  the BS  wave
function in $J^{P}=1^+$ channel, which is \mbox{$^3S_1-^3D_1$} coupled
system, we make a spatial average  with respect to the cubic group $O$
as  $\psi_{\alpha\beta}(\vec r;  ^3S_1)\equiv  \frac1{24}\sum_{g\in O}
\psi_{\alpha\beta}(g \vec r)$.
%%%...................................................................
Calculations are fully performed for $r \le 0.7$ fm, while, for $r \ge
0.7$ fm,  we restrict ourselves to  the points on  the coordinate axes
and  their nearest  neighbors to  reduce the  calculational  cost.  (A
rapid change of the potential is  expected for $r \le 0.7$ fm, whereas
a rather  mild change is  expected for $r  \ge 0.7$ fm.)  We  see that
there are  shrinks at short  distance, which suggest the  existence of
repulsion.
By using \Eq{eq.leading.central.1s0}, the central potential $V_{C}(r)$
in $^1S_0$ channel is constructed  at the leading order. The result is
shown in \Fig{fig.s-wave}(b).

To obtain the central potential in $^3S_1$ channel, it is necessary to
consider  a coupled  Schr\"odinger  equations of  $^3S_1$ and  $^3D_1$
channels, which will  be discussed in \Sect{section.tensor.potential}.
Here, we simply apply  the same formula \Eq{eq.leading.central.1s0} to
the wave function in $^3S_1$ channel.  Note that the resulting central
potential can  reproduce the  $^3S_1$ wave function  without involving
the tensor  potential.  Such  a central potential  is referred  as the
``{\em  effective central  potential}'', in  which the  effect  of the
tensor potential is embedded implicitly.   The result is also shown in
\Fig{fig.s-wave}(b).

In \Fig{fig.s-wave}(b), we wee that phenomenological properties of the
central nuclear potentials are  reproduced.  A repulsive core at short
distance  is surrounded  by  an attraction  at  medium distance.   The
effective  central  potential in  $^3S_1$  channel  tends  to be  more
attractive  than the central  potential in  $^1S_0$ channel.   This is
desirable for  the existence  of a bound  state (deuteron)  in $^3S_1$
channel in reality. (No bound state exists in $^1S_0$ channel.)

The    non-relativistic    energy    $E\equiv   \vec    k^2/m_N$    in
\Eq{eq.leading.central.1s0}  is obtained by  making a  fit of  BS wave
functions with Green's function of Helmholtz equation defined as
\begin{equation}
  (\triangle + k^2) G(\vec x; k^2) = -\delta_L(\vec x),
\end{equation}
where   $\delta_L(\vec  x)   \equiv   \sum_{\vec  n\in   \mathbb{Z}^3}
\exp\left( 2\pi i  \vec n\cdot \vec x/L \right)$  denotes the periodic
delta  function in the  three dimensional  torus of  spatial extension
$L$.
%%%...................................................................
The fit  is performed in the region  $11a \alt r \alt  16a$, where the
interaction is  seen to  become negligible from  a plot  of $\triangle
\psi(\vec    x)/\psi(\vec   x)$   \cite{ishizuka}.     The   resulting
non-relativistic energies are quite  small, i.e., $E = -0.509(94)$ MeV
for $^1S_0$  channel and $E =  - 0.560(110)$ MeV  for $^3S_1$ channel.
%%%...................................................................
Note that negative $E$ does not necessarily mean a formation of a bound
state.  This  is because  two nucleons cannot  be separated  from each
other beyond the range of  interaction in a finite volume.  The result
is also  shown in  \Fig{fig.s-wave}(a), where Green's  functions along
the coordinate axis are plotted with solid lines.
 
\begin{figure}[h]
\begin{center}
  \includegraphics[height=0.48\textwidth,angle=-90]{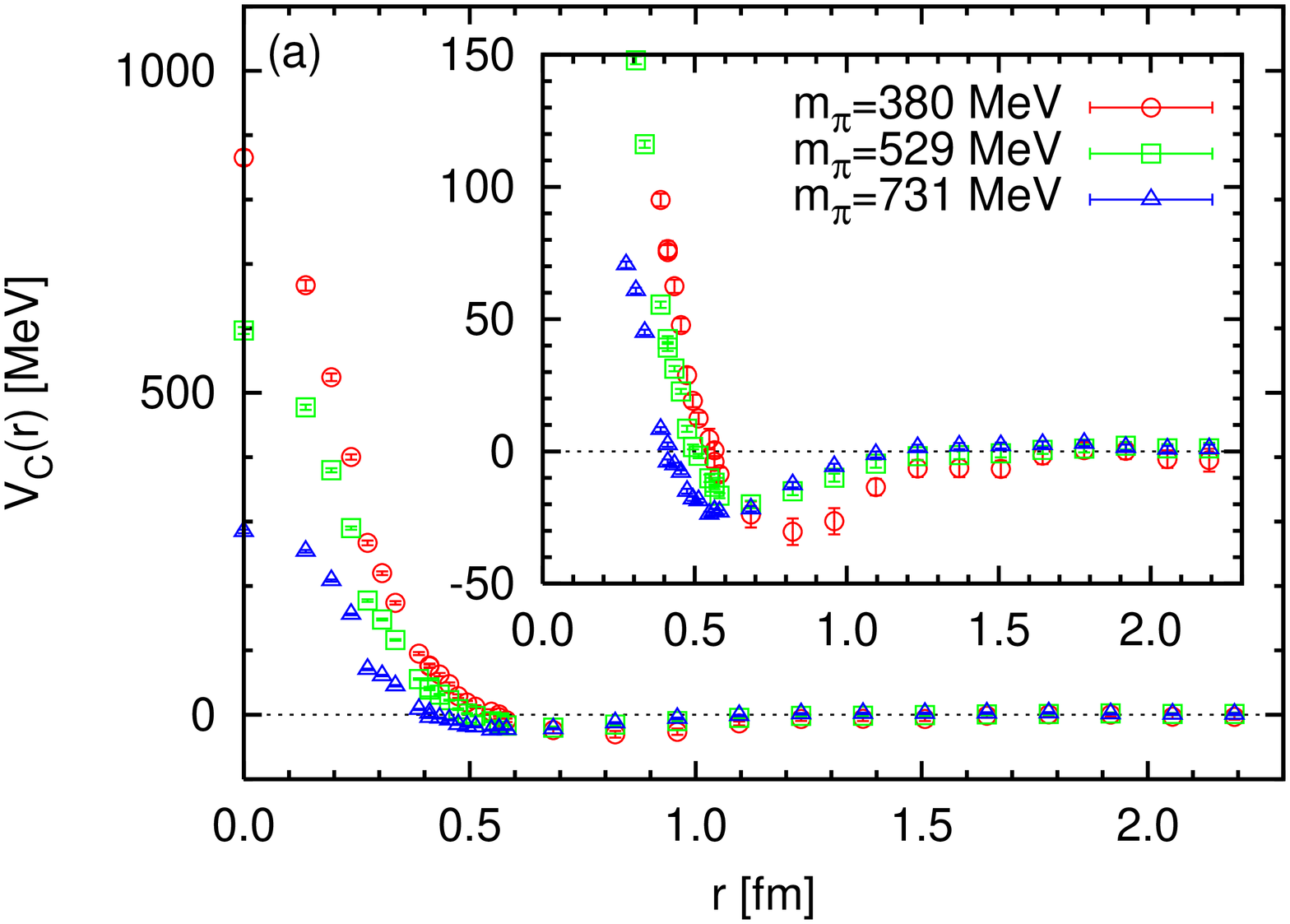}
  \includegraphics[height=0.48\textwidth,angle=-90]{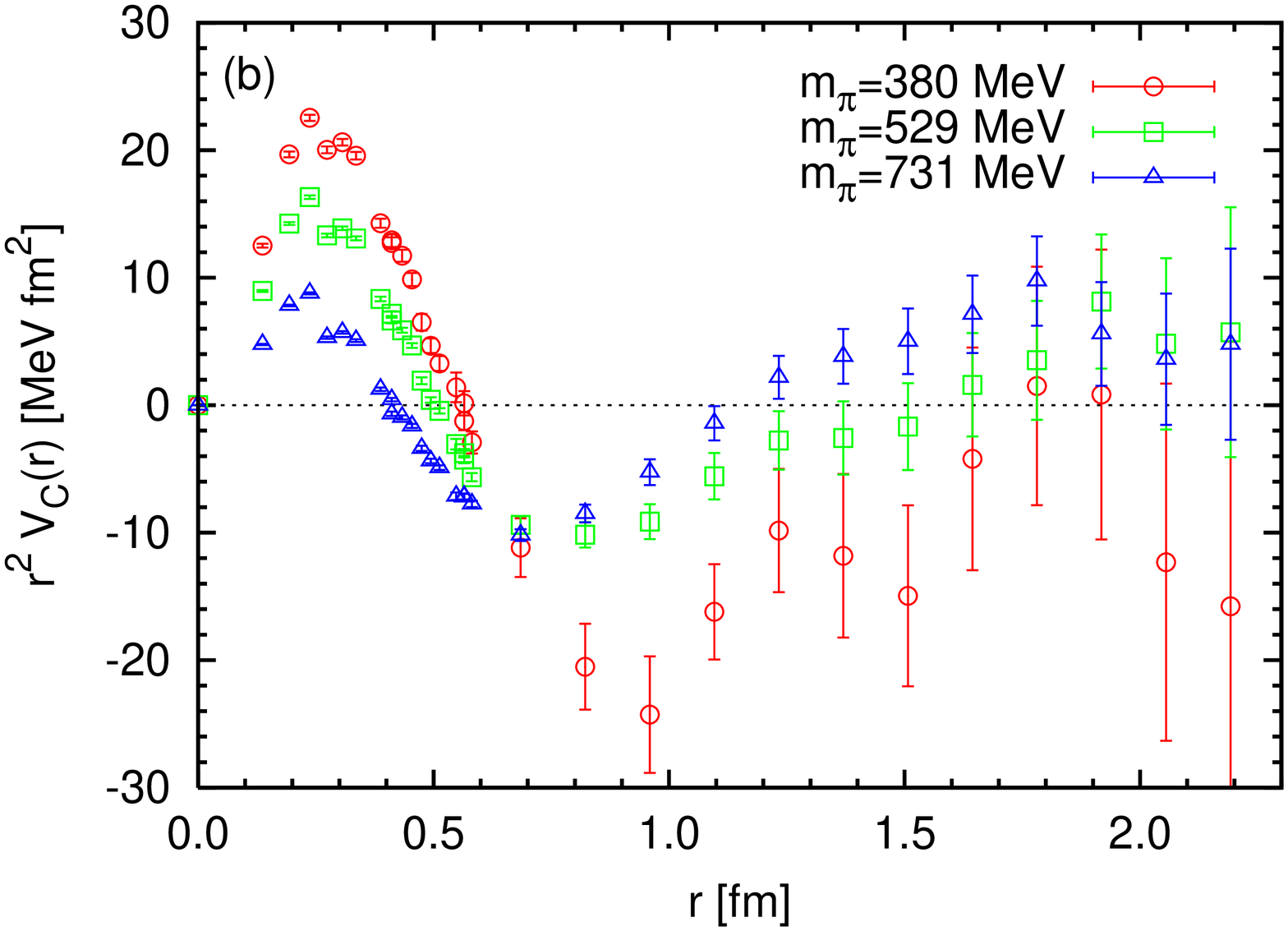}
\end{center}
\caption{(a) The central potential  in $^1S_0$ channel for three quark
masses and (b) those with $r^2$ multiplied. }
\label{fig.quark.mass.dependence}
\end{figure}
\Fig{fig.quark.mass.dependence}(a)  shows  the  central potentials  in
$^1S_0$ channel  at the  leading order for  three values of  the quark
mass.
%%%...................................................................
To see  the strength of the  potentials more accurately,  we plot them
with a  factor $r^2$ in  \Fig{fig.quark.mass.dependence}(b) to reflect
the effect of three-dimensional volume element.
We see that, as the quark  mass decreases, the repulsive core at short
distance  grows rapidly,  and  the attraction  at  medium distance  is
enhanced gradually.
The enhancement of the attraction is natural,
%%%...................................................................
since the smaller quark mass enables the virtual pion to propagate the
longer distance.
These   results  suggest   the  importance   of  direct   lattice  QCD
calculations in the light quark mass region.

\section{Tensor Potential at the leading order of the derivative expansion}
\label{section.tensor.potential}

In order  to construct  the central and  the tensor potentials,  it is
necessary  to take  into  account the  coupling  of s-wave($l=0$)  and
d-wave($l=2$)  components of  BS  wave function  in $J^P=1^+$  ($I=0$)
channel.
%%%...................................................................
Note  that the coupling  is induced  by the  tensor potential,  due to
which  the deuteron is  generated as  a bound  state in  reality.  The
tensor potential  plays a key role  in the stability  of atomic nuclei
and the saturation of nuclear  matter density. Its importance at short
distance  is pointed  out recently  by experimental  studies  of Short
Ranged Correlated (SRC) nucleon  pair, which may affects the structure
of the  cold dense  nuclear system such  as neutron  stars \cite{src}.
However, the  experimental determination of tensor  potential at short
distance is difficult, because it appears on top of the repulsive core
and the centrifugal barrier.

\begin{figure}[h]
\begin{center}
\includegraphics[height=0.48\textwidth,angle=-90]{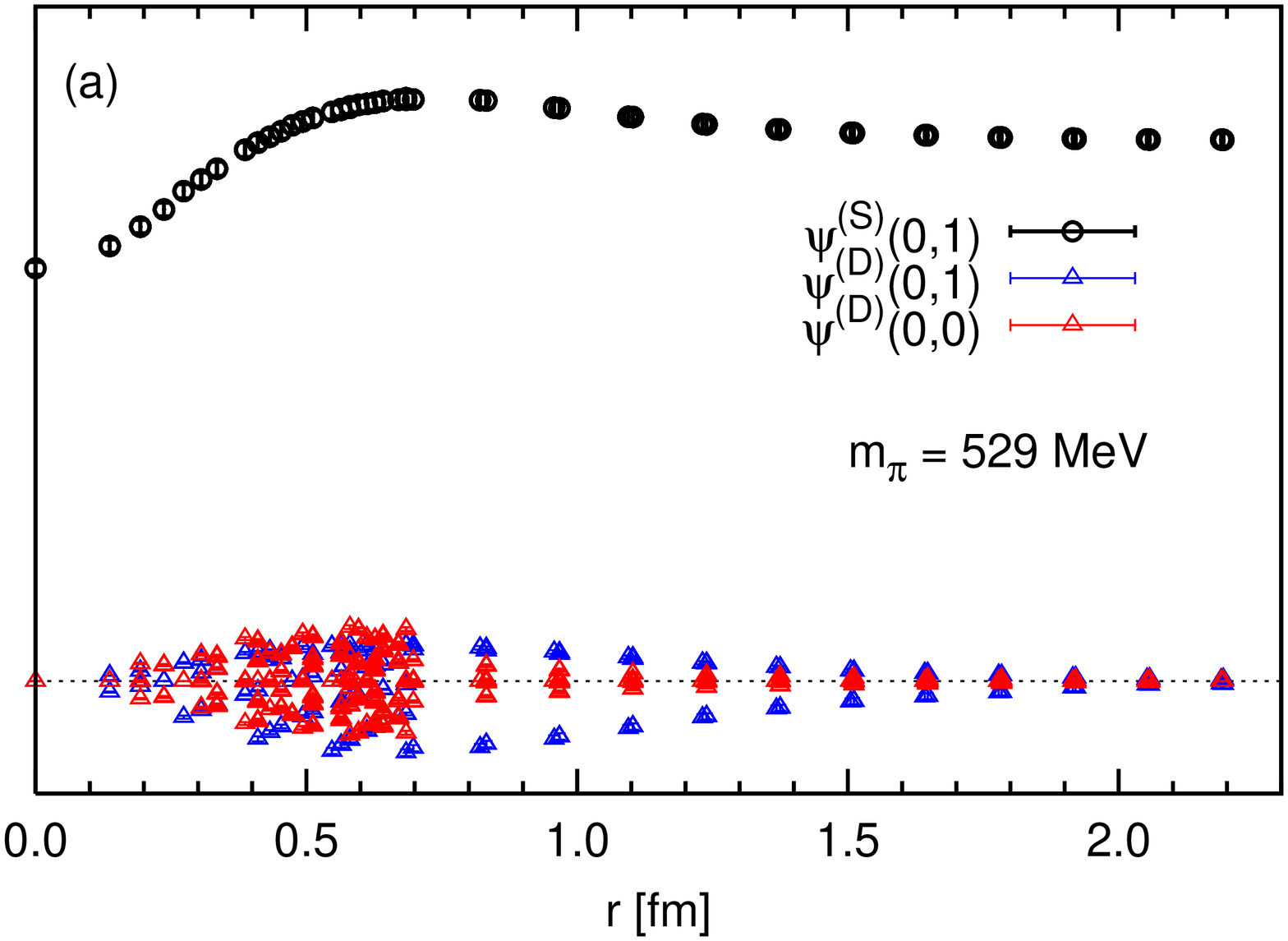}
\includegraphics[height=0.48\textwidth,angle=-90]{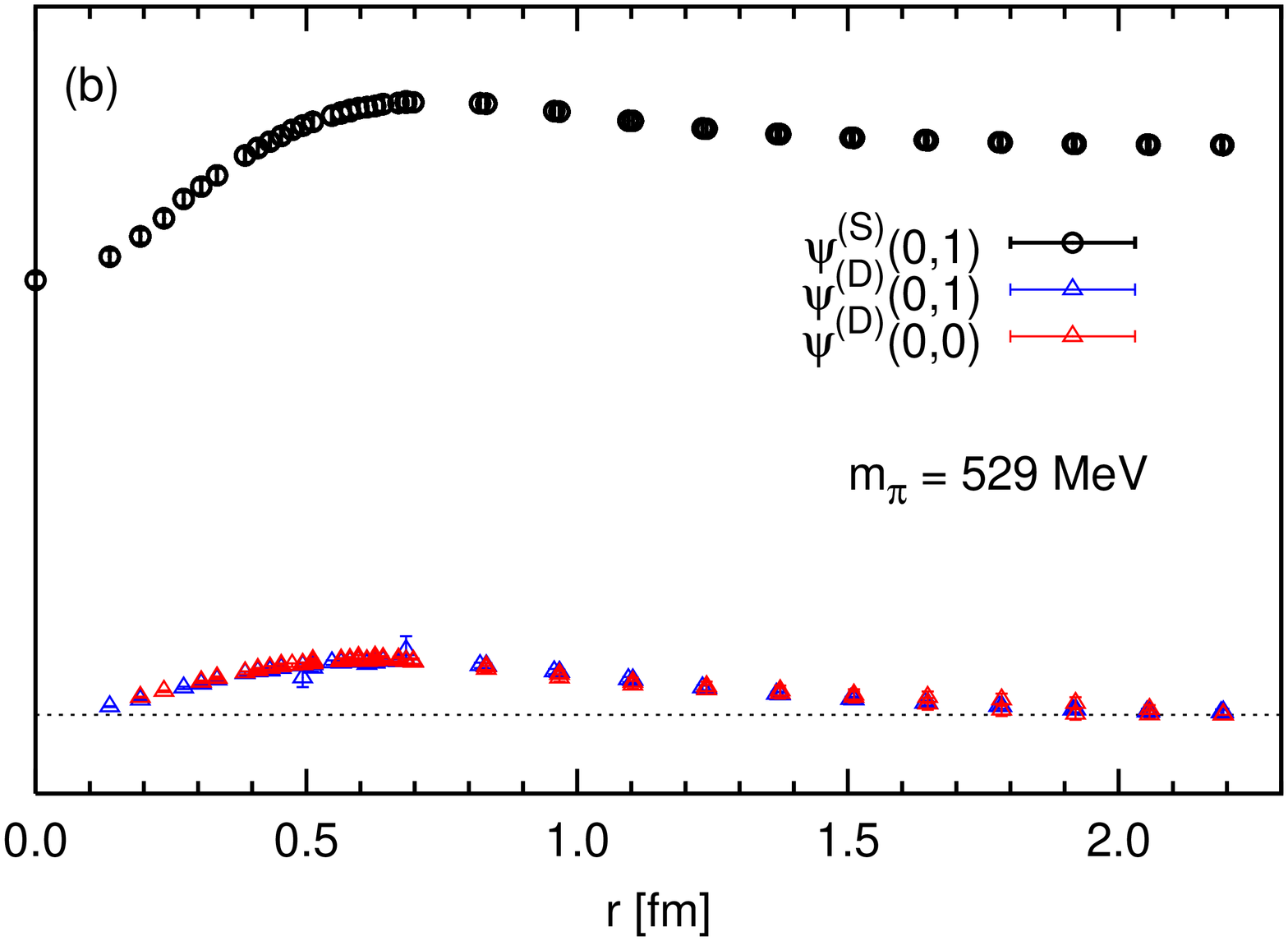}
\end{center}
\caption{(a) BS wave functions on the lattice and (b) BS wave function
after removing the spinor harmonics factors.}
\label{fig.d-wave}
\end{figure}

Construction  of tensor potential  depends on  the quality  of d-wave
wave function on the lattice.  On  the lattice, we construct a BS wave
function  in   $T_1^+$  representation  of  the   cubic  group,  which
corresponds to $J^P=1^+$ up to $J\ge 4$ contamination. We decompose it
into orbitally $A_1^+$  part $\psi^{({S})}(\vec r)$, which corresponds
to s-wave up to $l\ge 4$ contamination, and orbitally non-$A_1^+$ part
$\psi^{({D})}(\vec r)$.   The decomposition is performed  by using the
projection operators $P$ and $Q$ defined as
\begin{equation}
  \psi^{(S)}_{\alpha\beta}(\vec r)
  =
  P\left[\psi\right]_{\alpha\beta}(\vec r)
  \equiv
  \frac1{24}
  \sum_{g\in O}\psi_{\alpha\beta}(g \vec r),
  \Hs
  \psi^{(D)}_{\alpha\beta}(\vec r)
  =
  Q\left[\psi\right]_{\alpha\beta}(\vec r)
  \equiv
  \psi_{\alpha\beta}(\vec r) - \psi^{(S)}_{\alpha\beta}(\vec r),
\end{equation}
where  $O$  denotes  the  cubic  group with  24  elements.   From  the
orthogonality relations  of the characters of  the representations, we
find that the orbital part  of $\psi^{(D)}(\vec r)$ consists of either
$E^+$, $T_2^+$ or $T_1^+$ representations.
%%%...................................................................
$E^+$ and $T_2^+$  correspond to d-wave up to  $l\ge 4$ contamination.
Orbital $T_1^+$ representation corresponds  to g-wave ($l=4$) up to $l
\ge 6$,  which enters through  $J^P=4^+$ component of  $A_1^+$ through
the relation
$4^+(J^P) = 1(\mbox{spin})\otimes 4^+(\mbox{orbital})$.
%In  any   case,  we  convince  ourselves   that  $\psi^{(D)}(\vec  r)$
%
%We   consider    the   size    of   contamination   from    $l\ge   4$
%states.
%%%...................................................................
\Fig{fig.d-wave}(b)  shows BS wave  functions for  $T_1^+$ ($J^P\simeq
1^+$) and the azimuthal quantum number $M=0$.
%
%% {\bf   (For  $\psi^{(S)}$,   $(0,0)$  and   $(1,1)$   spin  components
%% vanish. $(0,1)$  and $(1,0)$  components are real,  and agree  to each
%% other.  For $\psi^{(D)}$, $(0,1)$ and $(0,1)$ components are real, and
%% agree  to each  other.   $(0,0)$ and  $(1,1)$  components are  complex
%% conjugate to each other.)}
%
We see that $\psi^{(S)}$  is single-valued, whereas $\psi^{(D)}$'s are
multi-valued.
%%%...................................................................
Since the angular dependence  manifests itself as multi-valuedness, it
follows  that $\psi^{(S)}$  is  dominated by  s-wave contribution.
To   consider  $\psi^{(D)}$,   we  note   that  d-wave   component  is
proportional  to the  ``{\em  spinor harmonics}''  in SO(3)  symmetric
limit as
\begin{equation}
  \left[
    \begin{array}{cc}
      \psi_{00}(\vec r) & \psi_{01}(\vec r) \\
      \psi_{10}(\vec r) & \psi_{11}(\vec r)
    \end{array}
    \right]
  \propto
  {\bf Y}_{J=1,M=0}^{(l=2)}(\hat r)
  \equiv
  \sqrt{3 \over 10}
  \times
  \left[
    \begin{array}{cc}
      Y_{2,-1}(\vec r) & -\frac{2}{\sqrt{6}} Y_{2,0}(\vec r) \\
      -\frac{2}{\sqrt{6}} Y_{2,0}(\vec r) & Y_{2,+1}(\vec r)
    \end{array}
    \right].
  \label{eq.spinor.harmonics}
\end{equation}
To  examine whether  $\psi^{(D)}$ is  dominated  by d-wave  or not,  we
divide      $\psi^{(D)}$      by      these      spinor      harmonics
factors.   \Fig{fig.d-wave}(b)  shows  the   results.   We   see  that
$\psi^{(D)}$ becomes single  valued, which indicates that $\psi^{(D)}$
is dominated by d-wave.

To separate the  s-wave and the d-wave parts,  we apply the projection
operators $P$ and $Q$ to \Eq{eq.leading.schrodinger}.
%%%...................................................................
Since $V_{C}(\vec r)$ and $V_{T}(\vec r)$ commute with $P$ and $Q$ due
to the rotational invariance, we have
\begin{eqnarray}
  -\frac{\triangle}{m_N}P\psi_{\vec k}(\vec r)
  + V_C(\vec r) P\psi_{\vec k}(\vec r)
  + V_T(\vec r) P S_{12}\psi_{\vec k}(\vec r)
  &=&
  \frac{k^2}{m_N} P \psi_{\vec k}(\vec r)
  \label{eq.separate.s.and.d}
  \\\nonumber
  -\frac{\triangle}{m_N}Q\psi_{\vec k}(\vec r)
  + V_C(\vec r) Q\psi_{\vec k}(\vec r)
  + V_T(\vec r) Q S_{12}\psi_{\vec k}(\vec r)
  &=&
  \frac{k^2}{m_N} Q \psi_{\vec k}(\vec r).
\end{eqnarray}
Note that  each of  these two equations  has two spinor  indices.  For
upper line, we have essentially a unique choice.  ($(0,1)$ and $(1,0)$
components agree to  each other up to an  overall sign.)  In contrast,
we can play with a particular  choice of spin components for the lower
line.   \Eq{eq.spinor.harmonics}  suggests  that $(0,1)$  and  $(1,0)$
component  correspond  to   $E$-representation,  whereas  $(0,0)$  and
$(1,1)$  component correspond  to  $T_2$-representation.  Since  SO(3)
symmetry is not exact, the results depend on how we choose d-wave wave
function.
%%%...................................................................
%% Although  these two  representations of  cubic group  correspond to
%% $l=2$, they also contain $l \ge 4$ component.
%
For simplicity in this section,  we choose $(0,1)$ spin component from
the lower line for d-wave.
%%%...................................................................
\Eq{eq.separate.s.and.d} is arranged as
\begin{equation}
  \left[
    \begin{array}{cc}
      P\psi_{\vec k}(\vec r), & PS_{12}\psi_{\vec k}(\vec r) \\
      Q\psi_{\vec k}(\vec r), & QS_{12}\psi_{\vec k}(\vec r)
    \end{array}
    \right]
  \cdot
  \left[
    \begin{array}{l}
      V_C(\vec r) - \frac{k^2}{m_N} \\
      V_T(\vec r)
    \end{array}
    \right]
  =
  \left[
    \begin{array}{c}
      \frac{\triangle}{m_N}
      P\psi_{\vec k}(\vec r)
      \\
      \frac{\triangle}{m_N}
      Q\psi_{\vec k}(\vec r)
    \end{array}
    \right],
\end{equation}
which can be algebraically  solved for $V_{C}(\vec r)$ and $V_{T}(\vec
r)$ point by point.
Unlike the central potential, the tensor potential does not involve an
additional  shift by  $-\frac{k^2}{m_N}$,  which adjusts  zero at  the
spatial infinity.
%%%...................................................................
%%%...................................................................
In \Sect{section.full.qcd},  we employ  another choice for  the d-wave
from the lower  line by combining the four  spinor components with the
spinor    harmonics   $\mbox{\bf    Y}(\hat   r)    \equiv   \mbox{\bf
Y}_{J=1,M=0}^{(l=2)}(\hat r)$ as
\begin{equation}
  \mbox{\bf Y}^{*}_{\alpha\beta}(\hat r)
  [Q\psi_{\vec k}]_{\alpha\beta}(\vec r)
  \cdot
  \left(
  V_C(\vec r) - \frac{k^2}{m_N}
  \right)
  +
  \mbox{\bf Y}^{*}_{\alpha\beta}(\hat r)
  [QS_{12}\psi_{\vec k}]_{\alpha\beta}(\vec r)
  \cdot
  V_{T}(\vec r)
  =
  \mbox{\bf Y}^{*}_{\alpha\beta}(\hat r)
  {\triangle \over m_N}
  [Q\psi_{\vec k}]_{\alpha\beta}(\vec r).
\end{equation}
Once  SO(3) is  realized as  a good  symmetry, this  becomes  the best
choice.
%%%...................................................................

\begin{figure}[h]
\begin{center}
  \includegraphics[height=0.48\textwidth,angle=-90]{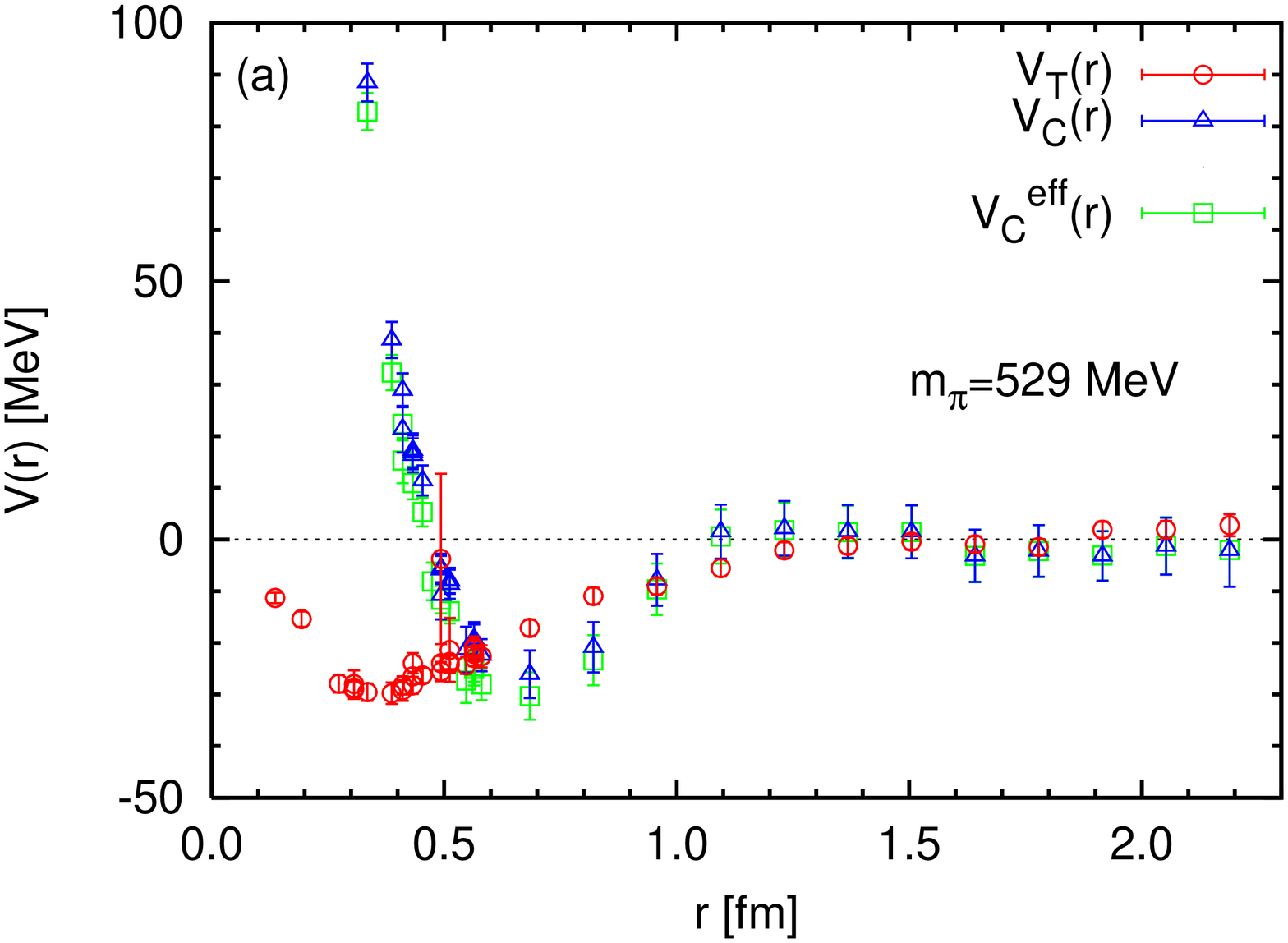}
  \includegraphics[height=0.48\textwidth,angle=-90]{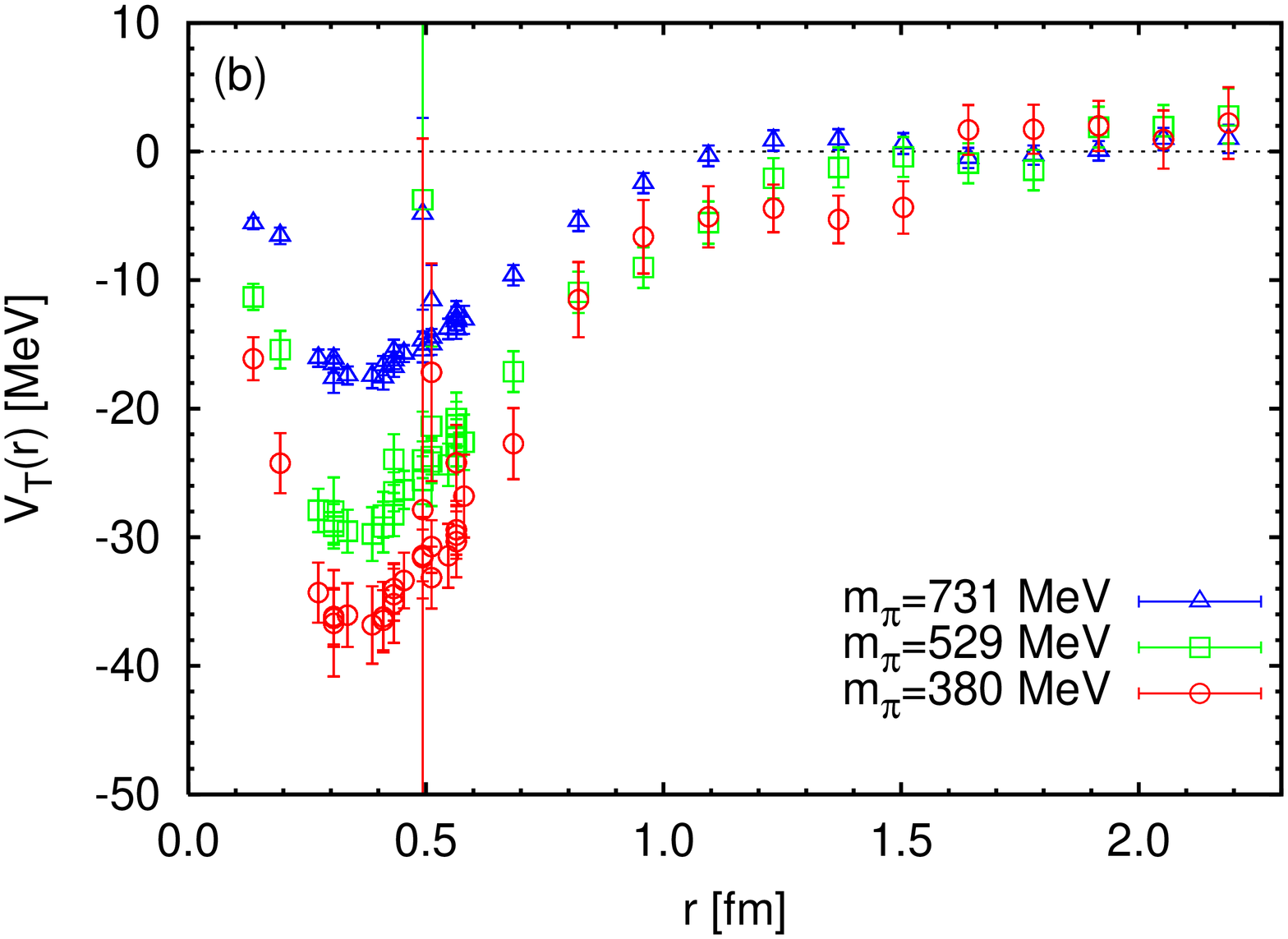}
\end{center}
\caption{  (a)  The tensor,  the  central  and  the effective  central
  potentials in $^3S_1-^3D_1$ coupled channel for $m_{pi}=529$ MeV and
  (b) the tensor potentials for three values of quark masses.}
\label{fig.tensor.force}
\end{figure}
\Fig{fig.tensor.force}(a)  shows the results  of the  tensor potential
together  with the  central and  the effective  central  potentials.
%%%...................................................................
%%% in $^3S_1-^3D_1$ coupled channel.
%%%...................................................................
The shape of our tensor potential is similar to the one-boson exchange
result,  which  is  obtained  by  the cancellation  between  the  pion
exchange and the rho meson exchange \cite{machleidt}.
%%%...................................................................
The difference between the central and the effective central potentials
is understood  by treating  the tensor potential  in the  second order
perturbation theory.   It seems to be  smaller than phenomenologically
expected. This is due to the  heavy quark mass.  Indeed, as is seen in
\Fig{fig.tensor.force}(b),  the tensor potential  is enhanced  in the
light quark mass region.

%%%...................................................................
A spike in the tensor potential at $r\simeq 0.5$ fm is due to the zero
of   $Y_{2,0}(\hat  r)   \propto   3\cos^2\theta  -   1$.  Note   that
$Y_{2,0}(\hat r)$  vanishes on  the lines $\vec  r = (\pm  n,\pm n,\pm
n)$.  In  the vicinity of these  lines, it becomes  difficult to solve
the  coupled  equation  numerically,  which leads  to  the  accumulated
statistical error.

\section{Convergence of the derivative expansion}

We examine  the convergence of  derivative expansion by  comparing two
potentials generated at two energies  $E\simeq 0$ MeV and $E\simeq 45$
MeV \cite{murano}.
%%%...................................................................
As soon  as the derivative  expansion is truncated,  energy dependence
may appear in the potential, because  it is not in general possible to
reproduce all the BS wave functions simultaneously only with truncated
degrees of freedom.
%%%...................................................................
In this way, the energy dependence  of the potential is related to the
non-locality of  the potential, which  makes it possible to  check the
convergence   of  derivative   expansion  by   examining   the  energy
dependence.

We  generate two  potentials  by imposing  different spatial  boundary
conditions on quark fields.
%%%...................................................................
A potential at $E\simeq 0$ MeV is generated with the periodic boundary
condition (PBC).   A potential at  $E\simeq 45$ MeV is  generated with
the anti-periodic  boundary condition  (APBC).  With APBC,  since each
nucleon consists of odd number of quarks, a nucleon is also subject to
APBC, so  that its spatial  momentum is discretized  as $p_i =  (2n_i +
1)\pi/L$ with  $n_i \in  \mathbb{Z}$.  For a  two nucleon  system, the
interaction  (nuclear  force) induces  a  modification  from its  free
value, i.e.,  $p^{(rel)}_i \simeq  (2 n_i +  1)\pi/L$.  Note  that the
smallest  spatial   momentum  for  APBC  is   $\vec  p^{(rel)}  \simeq
(\pm\pi/L,  \pm\pi/L, \pm\pi/L)$, which  does not  vanish. In  the box
with $L\simeq  4.4$ fm, $|\vec p^{(rel)}| \simeq  \sqrt{3}\pi/L = 244$
MeV.  Comparison  is made by  using the setup with  $m_{\pi}=529$ MeV,
$m_{N}\simeq 1333$ MeV, which  leads to the non-relativistic energy of
the lowest-lying state $E \equiv k^2/m_N \simeq 45$ MeV.
\begin{figure}[h]
\begin{center}
\includegraphics[width=0.32\textwidth]{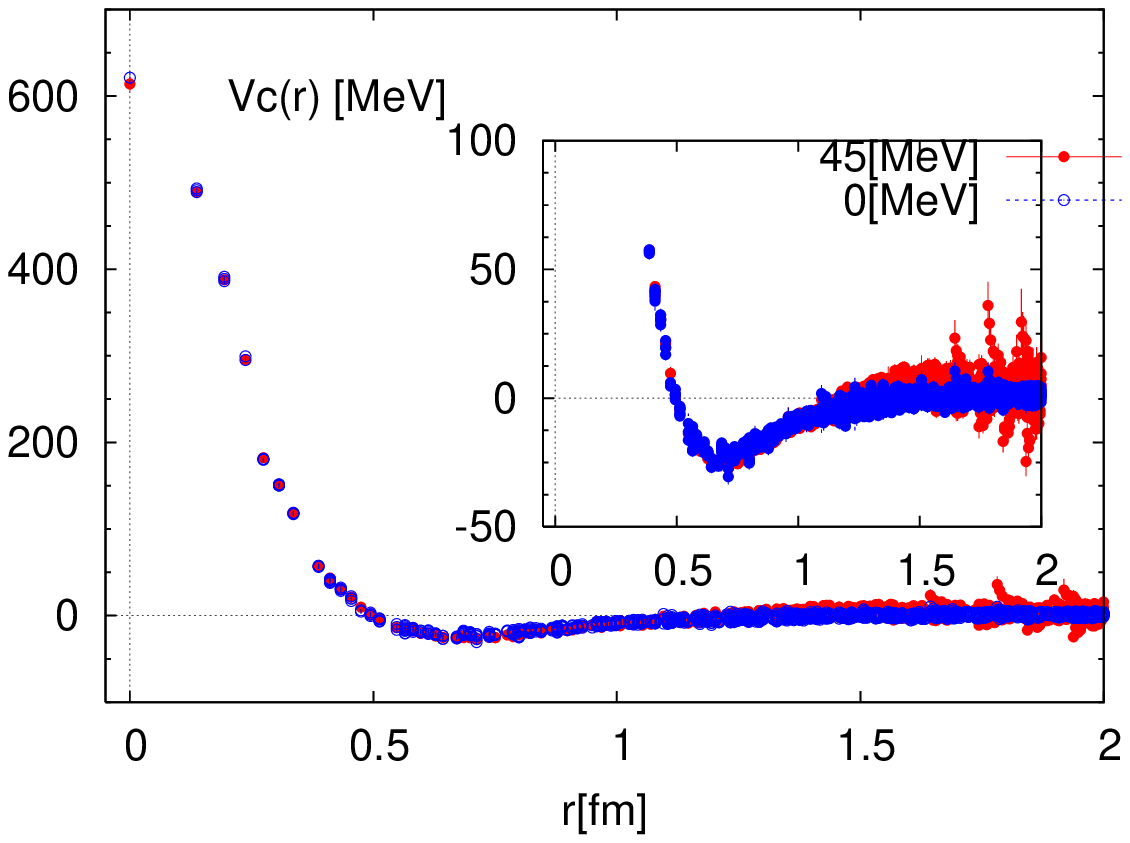}
\includegraphics[width=0.32\textwidth]{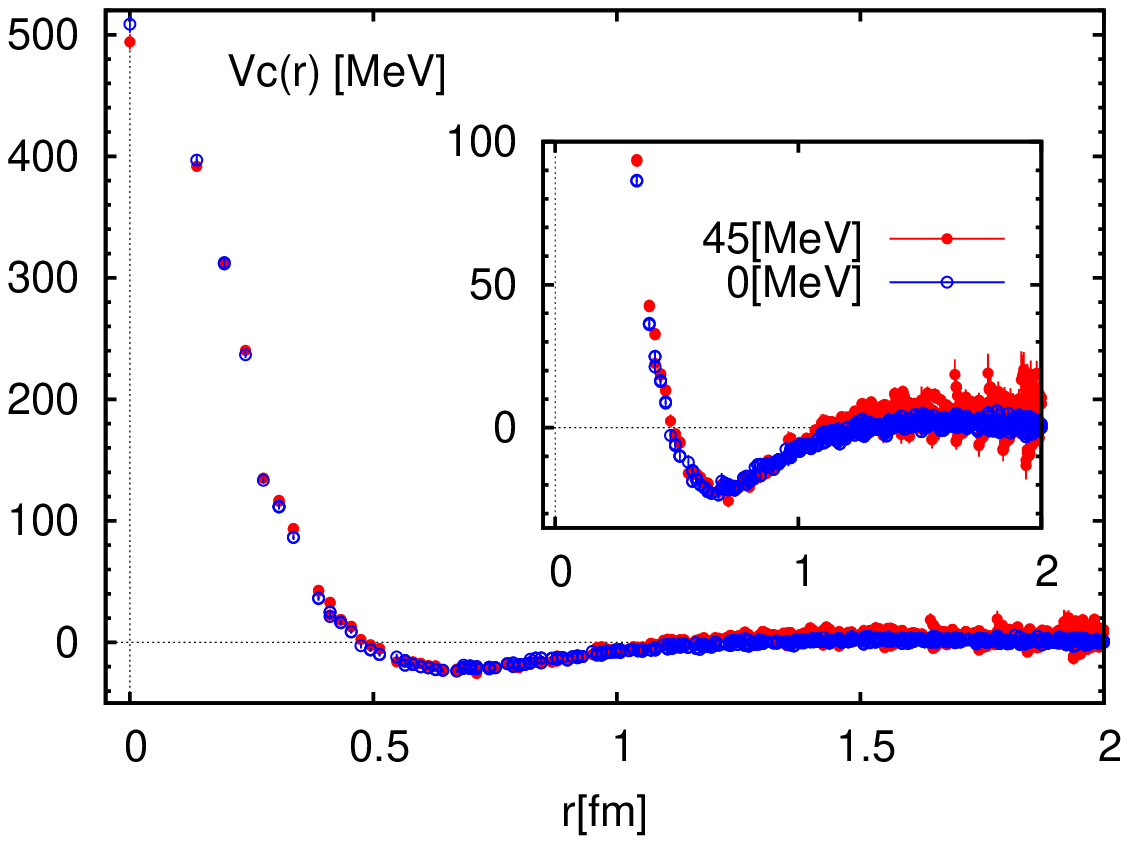}
\includegraphics[width=0.32\textwidth]{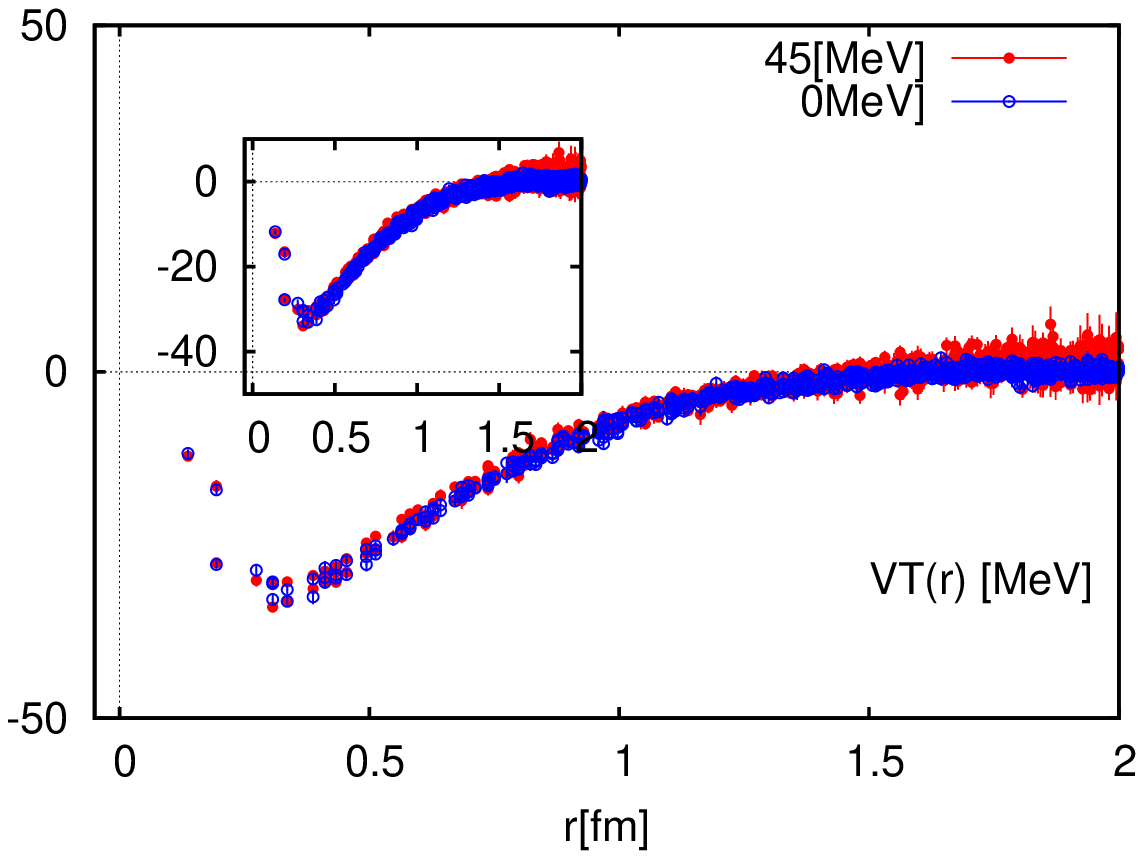}
\end{center}
\caption{Comparisons  of  leading order  potentials  generated at  two
  energies, i.e.,  $E\simeq 0$ MeV by  PBC and at $E\simeq  45$ MeV by
  APBC. The left, middle, right  figures show the central potential in
  $^1S_0$  channel,   the  central   and  the  tensor   potentials  in
  $^3S_1-^3D_1$ coupled channel, respectively.}
\label{fig.pbc.vs.apbc}
\end{figure}
The  results  are shown  in  \Fig{fig.pbc.vs.apbc}.  We  see that  the
agreement is quite good except small deviations at short distance.
The structures appearing in the region $r \agt 1$ fm for APBC turn out
to  be   caused  by  a   small  contamination  of  an   excited  state
\cite{murano}.
It follows  that the  derivative expansion works,  and that  our local
potential constructed at the leading order can be reliably used in the
energy region $E\simeq 0 - 45$ MeV.  (See \Ref{murano} for detail.)

\section{2+1 flavor QCD result of nuclear force
  with PACS-CS gauge configuration}  
\label{section.full.qcd}

In order to study the  quantitative features of nuclear potentials, it
is necessary to  resort to dynamical QCD performed  in the light quark
mass region employing a large spatial volume.
%%%...................................................................
PACS-CS Collaboration  is generating such  gauge configurations, i.e.,
2+1 flavor  gauge configurations, which cover the  physical quark mass
employing large spatial volumes $L\sim 3-6$ fm \cite{pacscs}.
%%%...................................................................
%% We use  some of  these gauge configurations  to obtain 2+1  flavor QCD
%% results of nuclear potentials.
%%%...................................................................
We use PACS-CS  gauge configurations to obtain 2+1  flavor QCD results
of  nuclear  potentials. The  gauge  configurations  are generated  by
employing  Iwasaki gauge  action  at $\beta=1.90$  on $32^3\times  64$
lattice  and  O($a$)-improved  Wilson  quark (clover)  action  with  a
non-perturbatively   improved  coefficient   $c_{\rm  SW}   =  1.715$.
$m_{\pi}$, $m_{K}$  and $m_{\Omega}$ are  used to determine  the scale
unit  $a^{-1} =  2.176(31)$ GeV  ($a\simeq 0.091$  fm) leading  to the
spatial  extension  $L =  32  a  \simeq  2.90$ fm  \cite{pacscs}.   To
calculate  nuclear potentials, we  use three  series of  PACS-CS gauge
configurations  with $(\kappa_{ud},  \kappa_{s})  = (0.16700,0.16400),
(0.16727,0.16400)$   and  $(0.16754,0.16400)$,  which   correspond  to
$m_{\pi}  \simeq 701$,  $570$ and  $411$ MeV  and $m_{N}  \simeq 1583,
1412$ and $1215$ MeV, respectively.

\begin{figure}[h]
\begin{center}
  \includegraphics[height=0.48\textwidth,angle=-90]{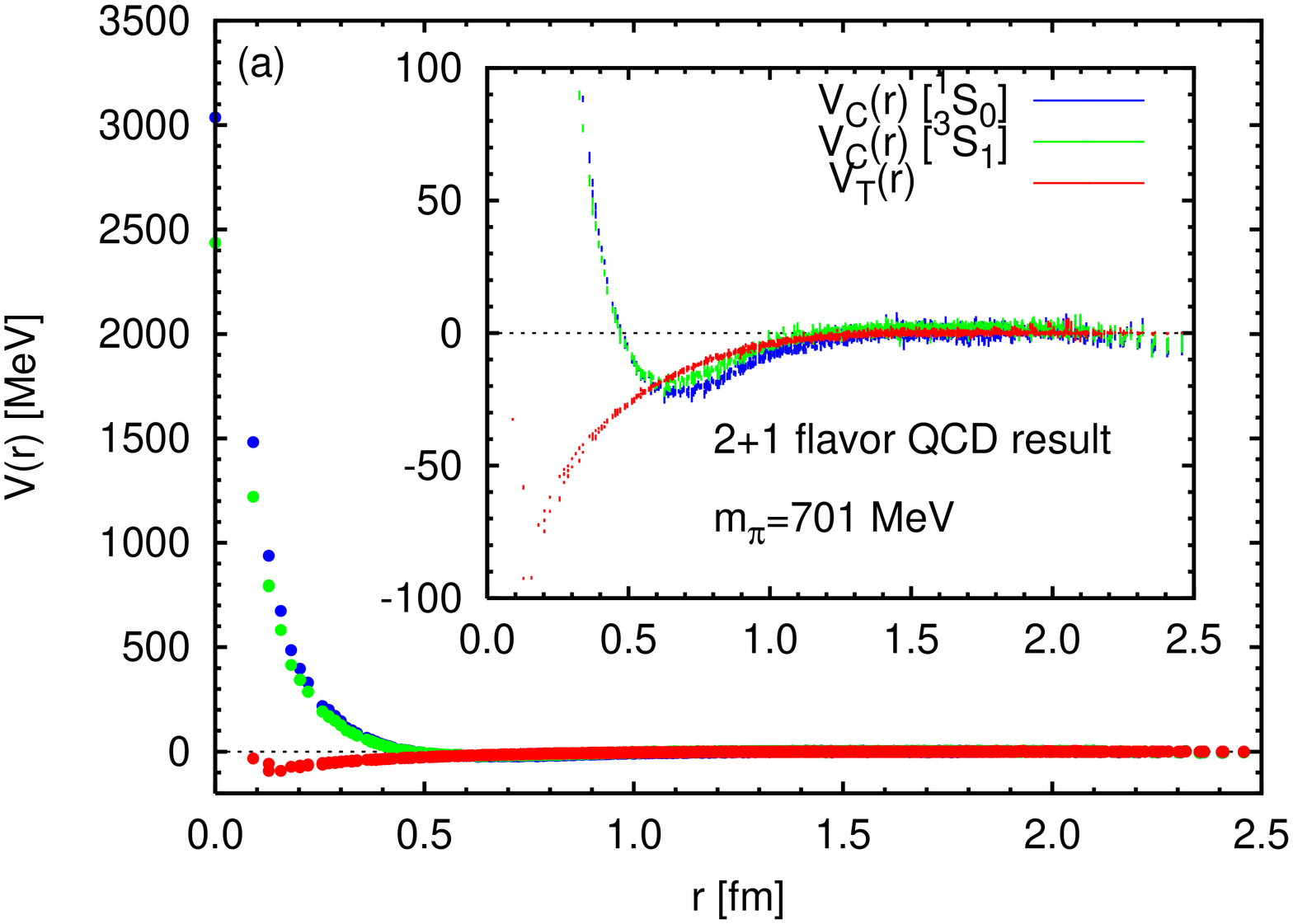}
  \includegraphics[height=0.48\textwidth,angle=-90]{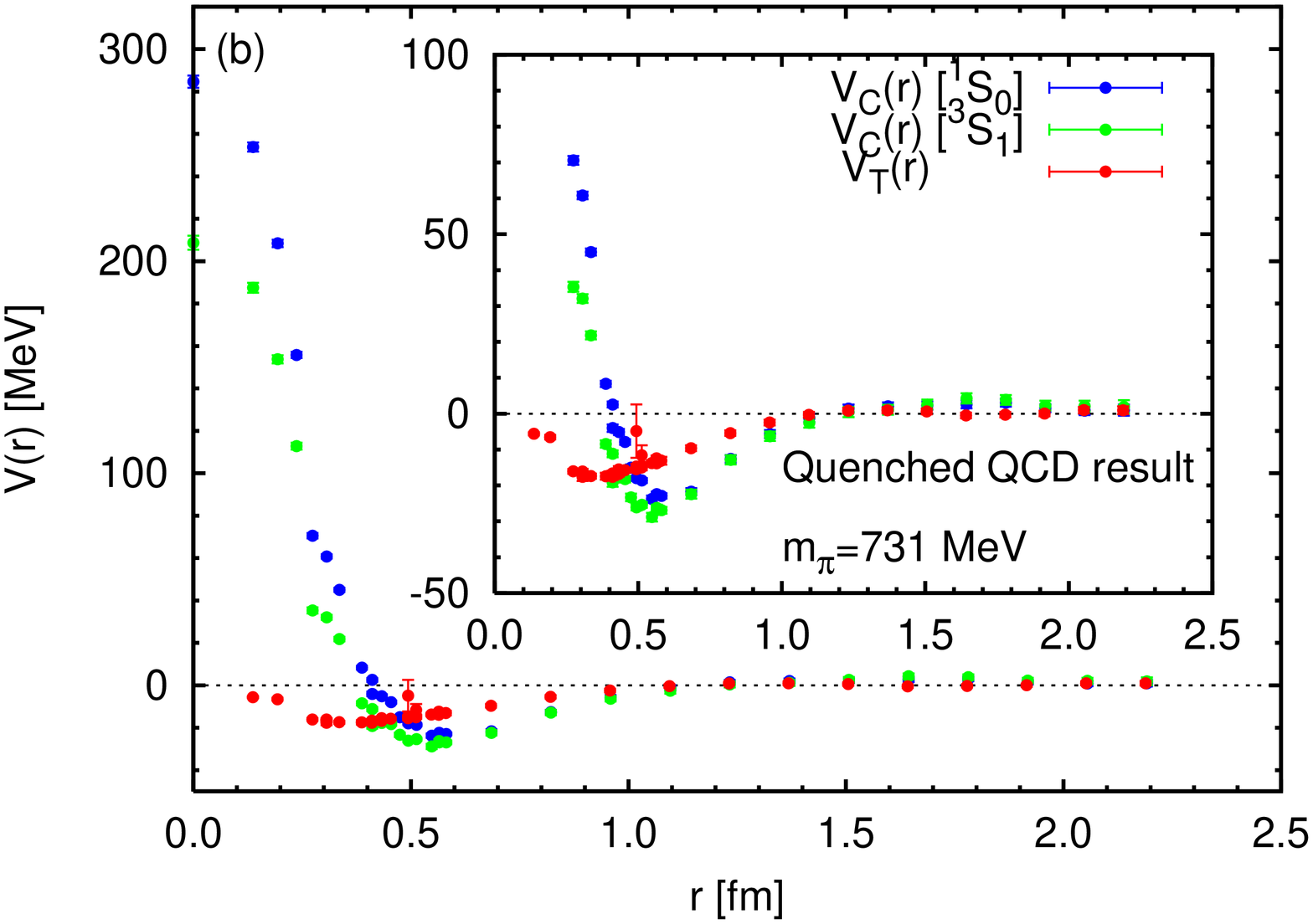}
\end{center}
\caption{(a) 2+1 flavor QCD results of the central and tensor potentials
  for $m_{\pi} = 701$ MeV,
  and (b) quenched QCD results of the central and tensor potentials
  for $m_{\pi} = 731$ MeV.}
\label{full.vs.quench}
\end{figure}
\Fig{full.vs.quench}(a)  shows  the  2+1  flavor QCD  results  of  the
nuclear  potentials for  $m_{\pi}  \simeq 701$  MeV,  which should  be
compared  with  the  quenched  results in  \Fig{full.vs.quench}(b)  of
comparable pion mass $m_{\pi} \simeq 731$ MeV.
%%%...................................................................
We  see that  the repulsive  cores at  short distance  and  the tensor
potential  become significantly  enhanced.  The  attraction  at medium
distance  tends to  be shifted  to  outer region,  whereas it  remains
almost unchanged in magnitude.
%%%...................................................................
Although these changes may be  caused by dynamical quarks, they may be
due to a lattice discretization artifact.
%%%...................................................................
We need further information to conclude.

\begin{figure}[h]
  \begin{center}
    \includegraphics[height=0.325\textwidth,angle=-90]{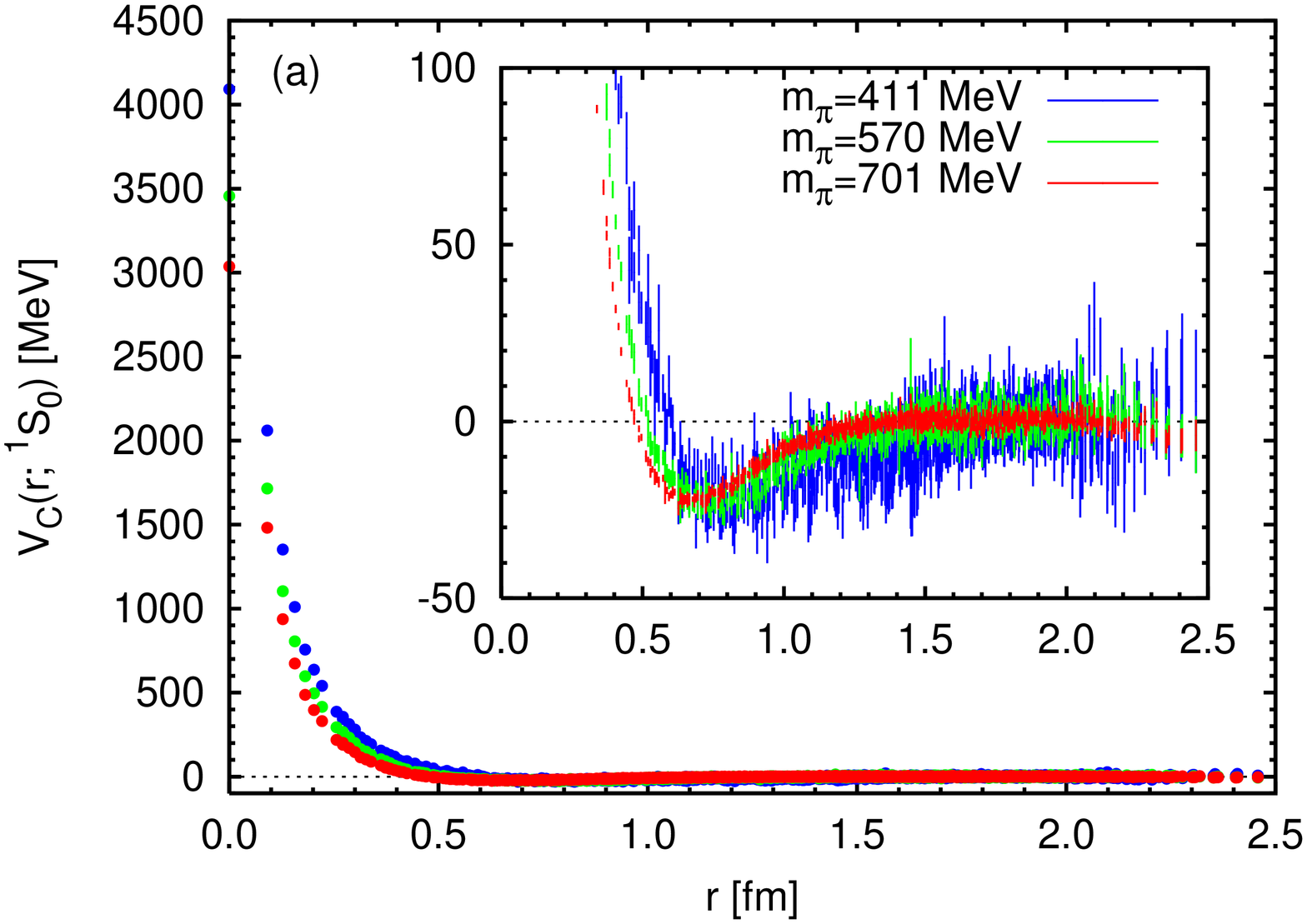}
    \includegraphics[height=0.325\textwidth,angle=-90]{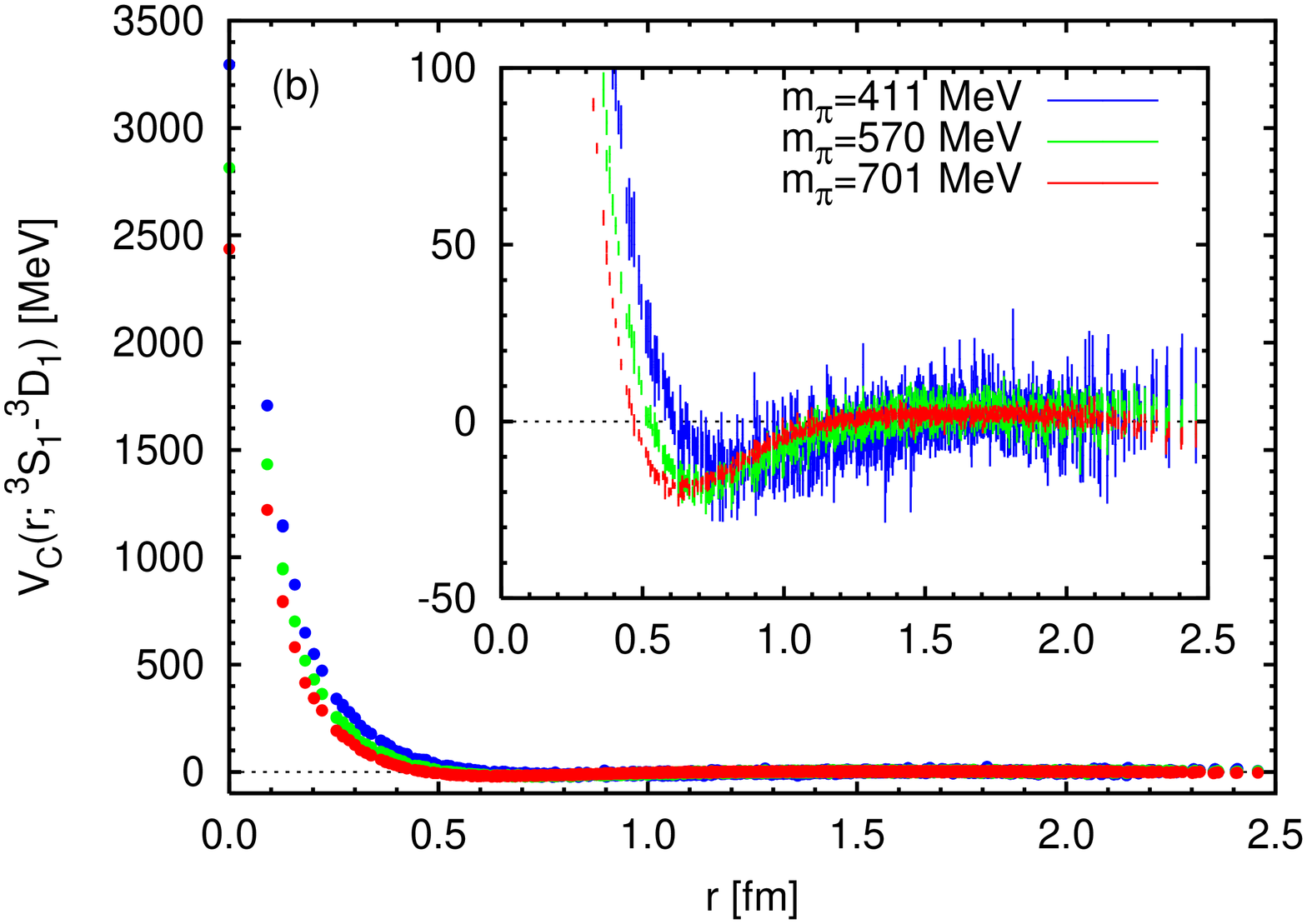}
    \includegraphics[height=0.325\textwidth,angle=-90]{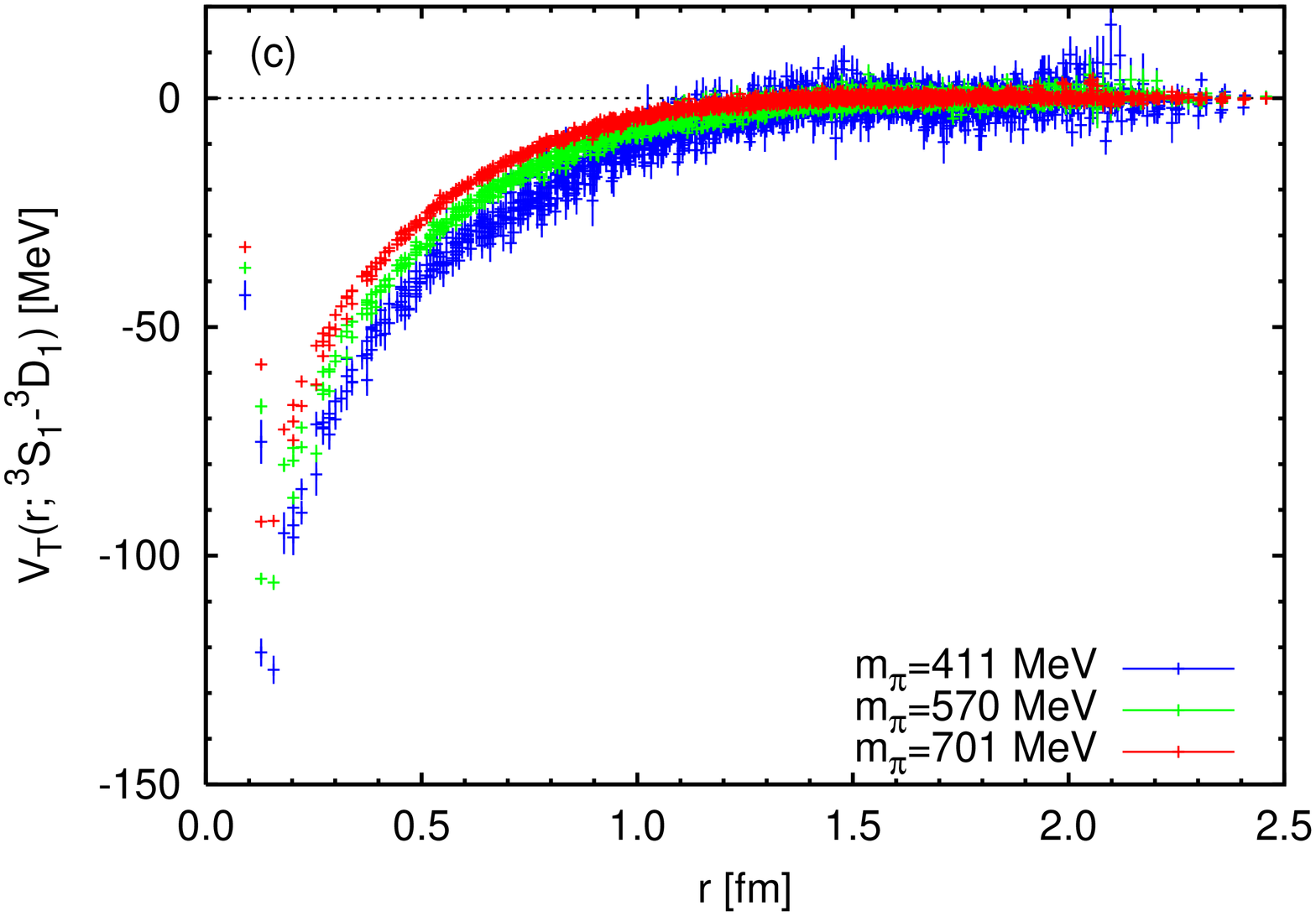}
    \\
    \includegraphics[height=0.325\textwidth,angle=-90]{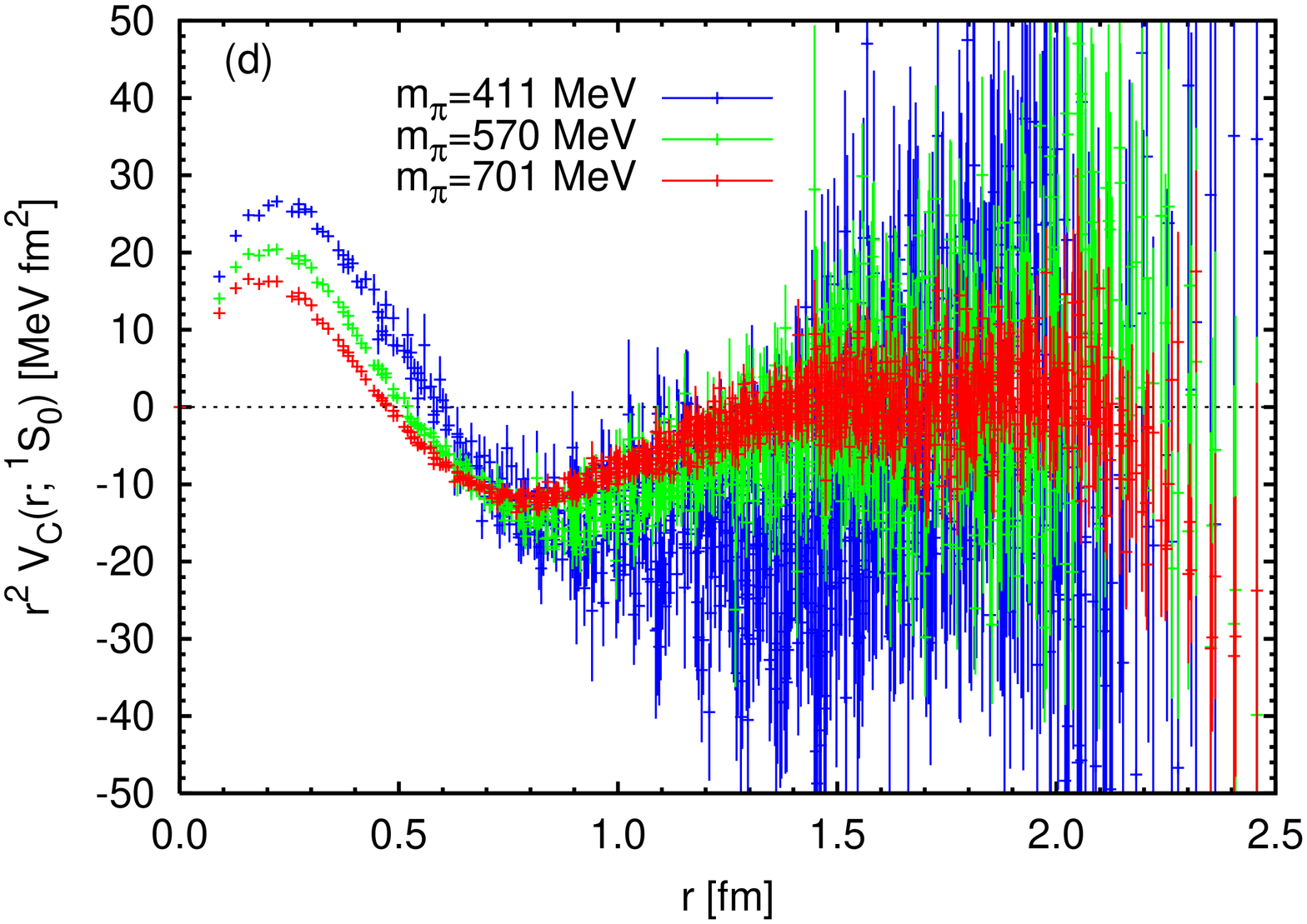}
    \includegraphics[height=0.325\textwidth,angle=-90]{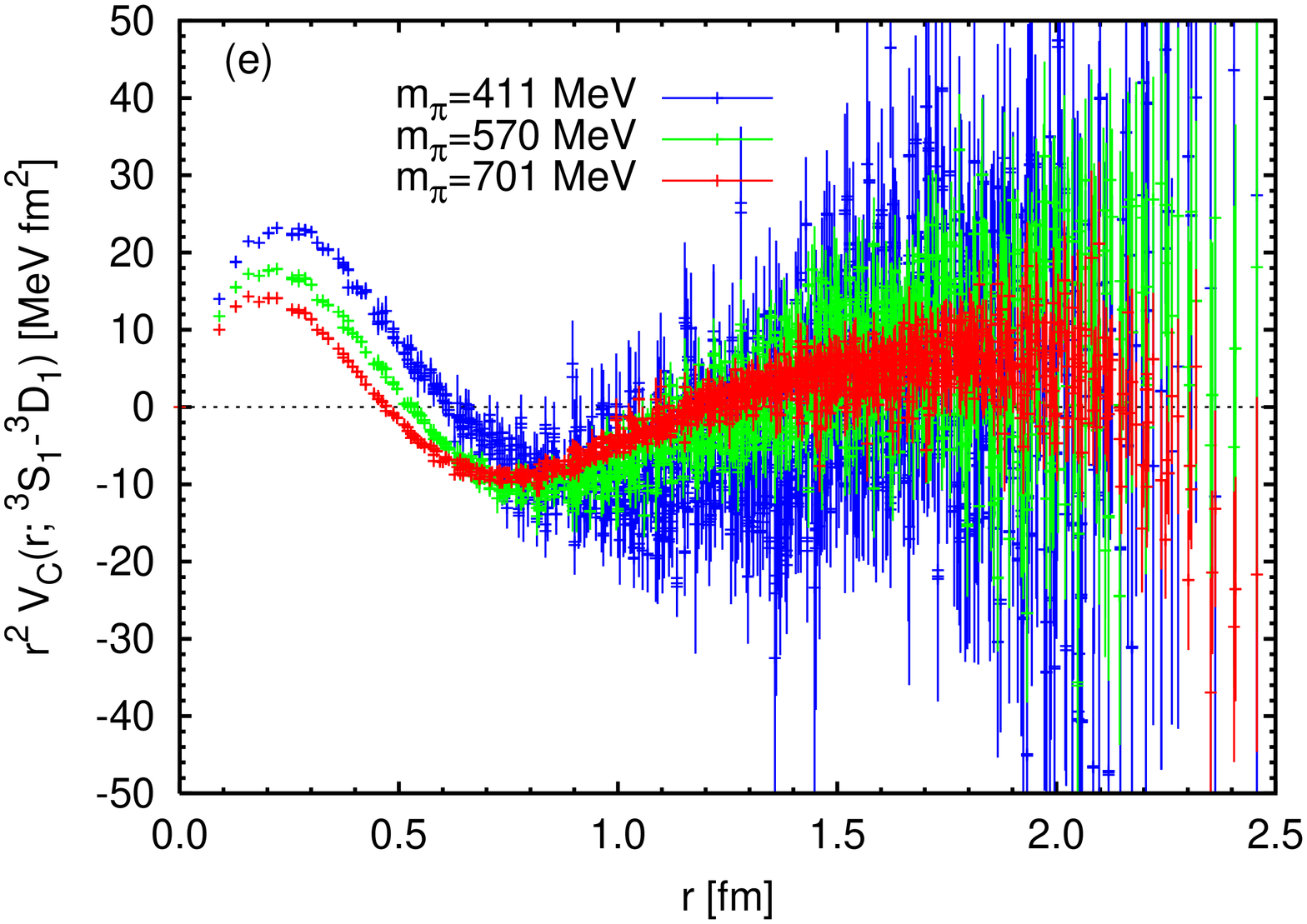}
    \includegraphics[height=0.325\textwidth,angle=-90]{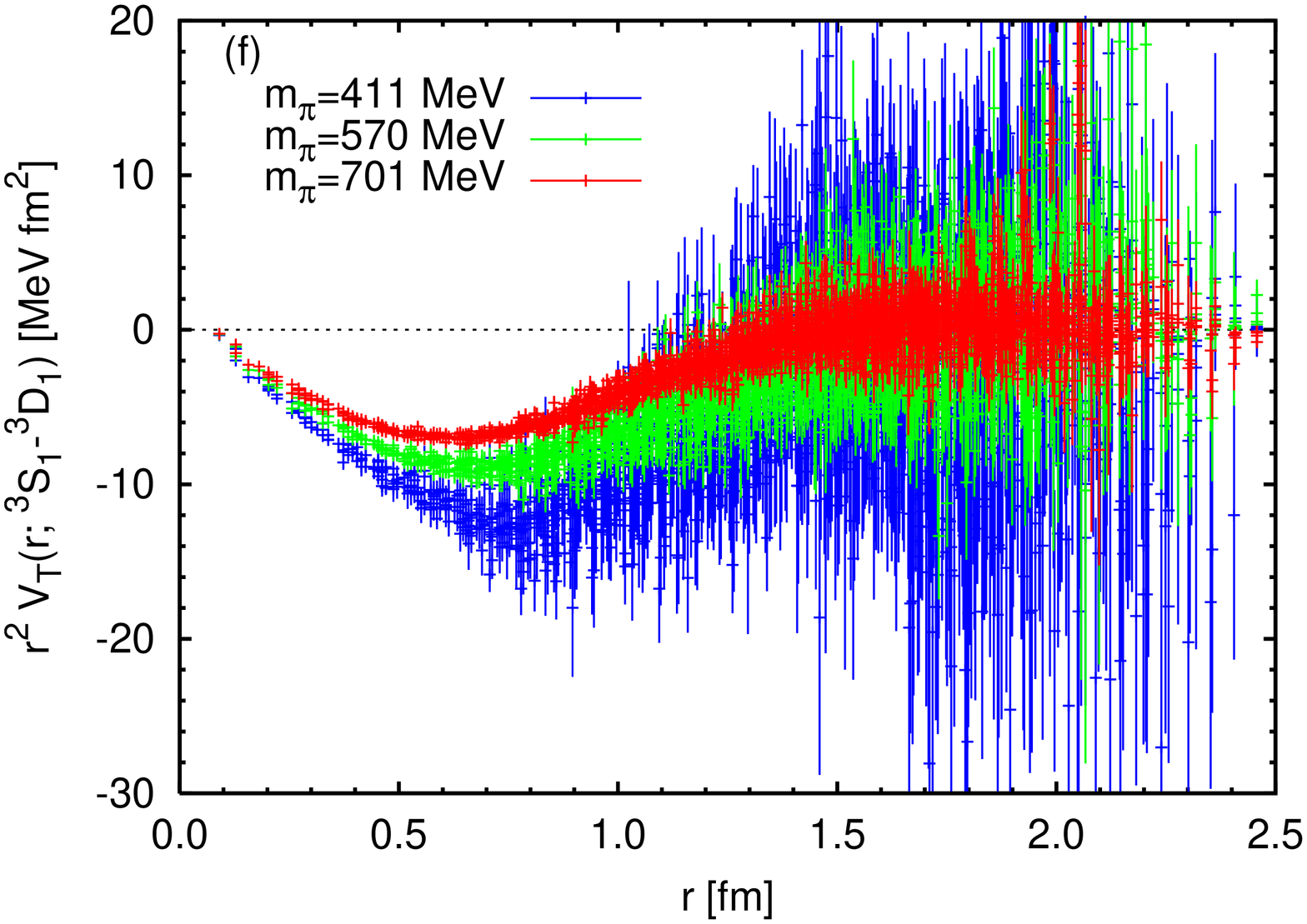}
  \end{center}
  \caption{$2+1$  flavor QCD  results  of nuclear  potentials for  three
    values  of quark  mass. (a)  $V_{\rm  C}(r)$ in  $^1S_0$ channel,  (b)
    $V_{\rm C}(r)$  in $^3S_1-^3D_1$  coupled channel, (c)  $V_{\rm T}(r)$
    for $^3S_1-^3D_1$ coupled channel,
    %%%...................................................................
    (d) $r^2V_{\rm  C}(r)$ in $^1S_0$  channel, (e) $r^2 V_{\rm  C}(r)$ in
    $^3S_1-^3D_1$   coupled   channel,   (f)   $r^2  V_{\rm   T}(r)$   for
    $^3S_1-^3D_1$ coupled channel.}
  \label{fig.full.1}
\end{figure}
Fig.~\ref{fig.full.1}  shows  the   central  potential  $V_{C}(r)$  in
$^1S_0$  channel and the  central potential  $V_{C}(r)$ and  the tensor
potential $V_{T}(r)$  in $^3S_1-^3D_1$ coupled  channel, together with
those  $r^2$  multiplied.   Similar  tendencies are  observed  as  the
quenched QCD, such as the enhancements of the repulsive cores at short
distance, the attractive pockets  at medium distance, and the strength
of the tensor potential.
%%%...................................................................
%% Because of  the limited spatial volume,  i.e., $L\simeq 3$  fm, we may
%% not go to a lighter quark mass region.

\begin{figure}[h]
\begin{center}
  \includegraphics[height=0.48\textwidth,angle=-90]{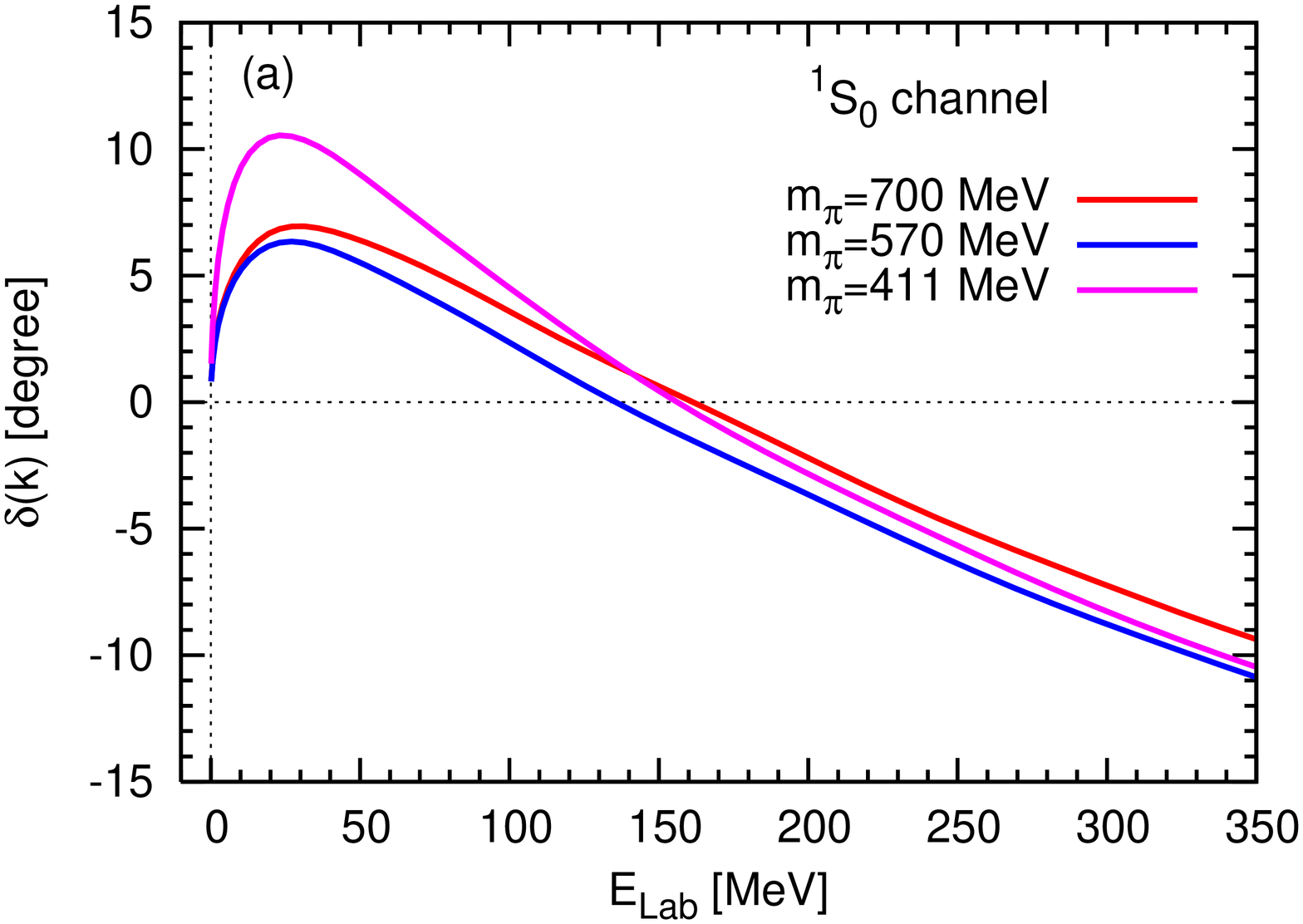}
  \includegraphics[height=0.48\textwidth,angle=-90]{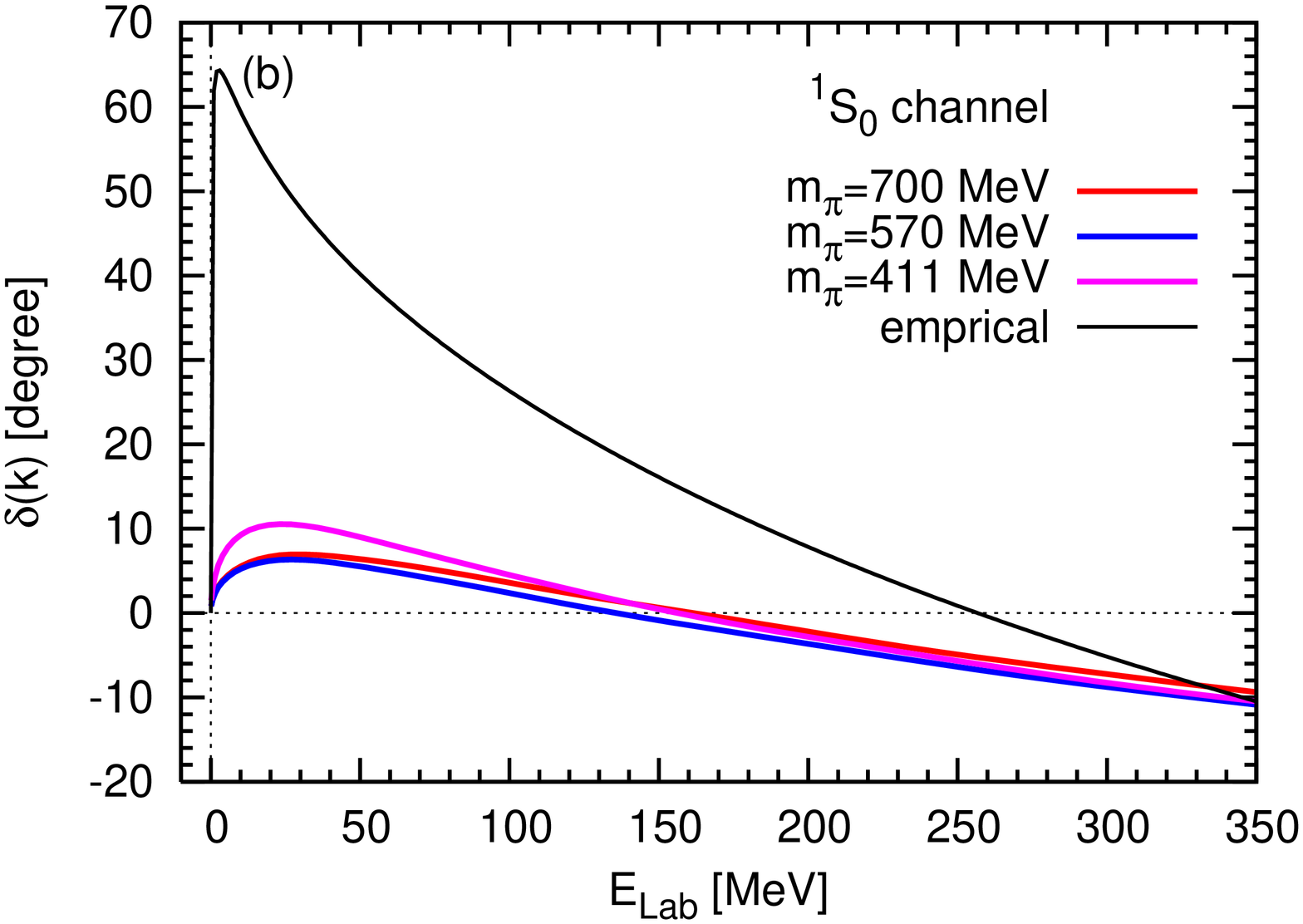}
\end{center}
\caption{(a)  Scattering  phase shifts  in  $^1S_0$  channel from  the
lattice nuclear potentials and (b) those with the empirical one.}
\label{fig.phase.shift}
\end{figure}
\Fig{fig.phase.shift} shows the phase shifts obtained from the nuclear
potentials  (not by  L\"uscher's method).   At low  energy,  the phase
shift grows up, which is  caused by the attraction at medium distance.
At high  energy, the phase shift  decrease, which is  a consequence of
the repulsive core at short distance.  One may wonder why the order of
the  phase shifts  of $m_{\pi}=570$  and $701$  MeV is  inverted.  The
reason seems to  be that the repulsive core grows  more rapid than the
attraction grows.  Qualitative shape of  the phase shift is seen to be
reasonable, which is due to  the fact that the qualitative features of
the nuclear potential are already reproduced.  However, comparing with
the  empirical one,  the strength  is  not satisfactory  at all,  which
suggests the importance of the light quark mass effect.

A technical comment is in order.
%%%...................................................................
To obtain the scattering length, the ground state saturation has to be
achieved to an accuracy of, at most, around 1 MeV, which is about 0.05
\% of the total mass of  the two nucleon system.  For the moment, such
a high precision is not yet attained in our calculation.
%%%...................................................................
Significantly  large $t$  and,  accordingly, the  large statitics  are
required.
%%%...................................................................
To calculate only the scattering  length, the smeared source is better
than  the wall  source for  such  high precision  calculations.
%%%...................................................................
The smeared  source, with  the help of  the average over  the relative
coordinate $\vec x$  in the sink side, projects  out the excited state
contamination from the temporal correlation.
%%%...................................................................
However, to calculate the BS wave function, which measures the spatial
correlation,  the sink  has to  be  unaveraged, and  the ground  state
saturation has to be achieved  point by point uniformly in the spatial
directions.
%%%...................................................................
Note  that the  smeared source  creates a  spatially squeezed  BS wave
function in the small $t$  region, which gradually broadens during the
temporal evolution,  until the ground  state shape is  achieved.  Here,
the convergence in the region $|\vec  x| \agt 1$ fm is quite slow, and
unreasonably large $t$  is required for the uniform  saturation by the
ground state.
%%%...................................................................
For future  applications to  the nuclear physics,  it is  necessary to
seek  for a  better source,  which makes  it possible  to  achieve the
uniform  saturation  of  the   ground  state  BS  wave  function  more
efficiently.

\section{Hyperon potentials}

\begin{figure}[h]
\begin{center}
\includegraphics[width=0.48\textwidth]{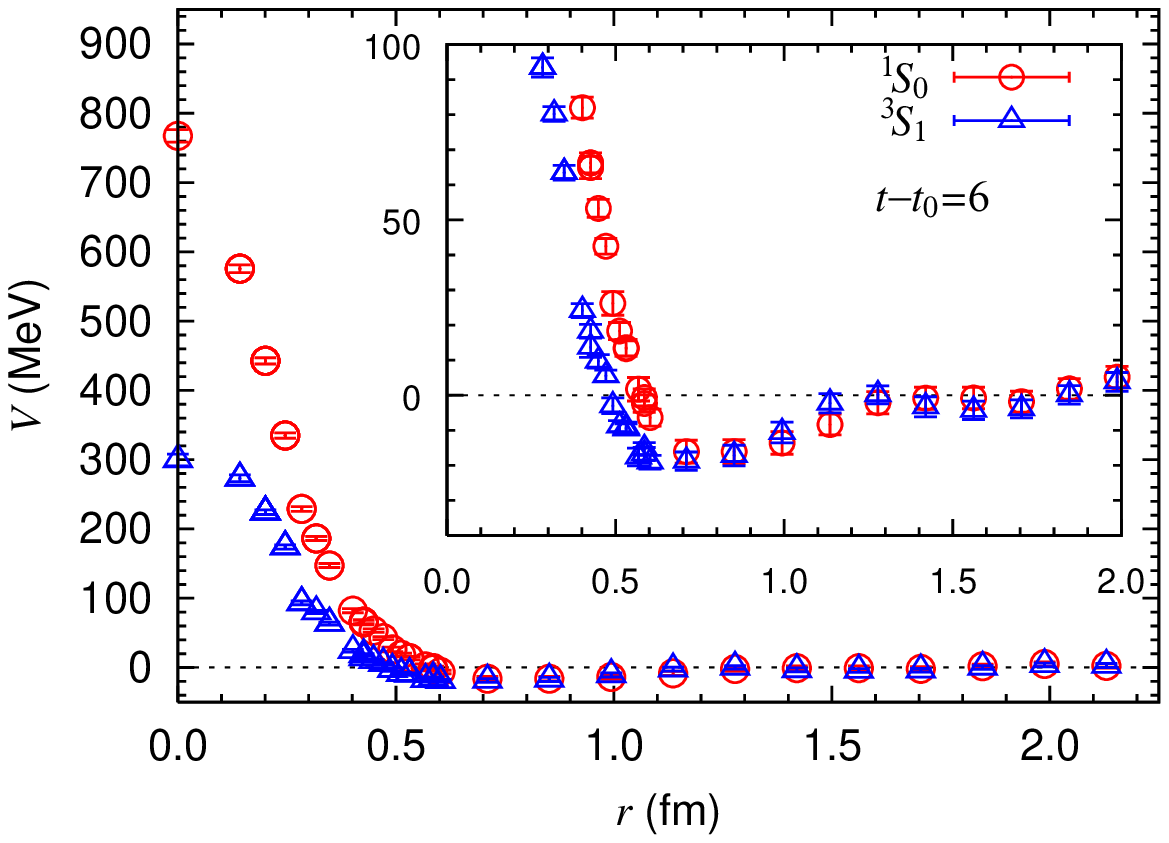}
\includegraphics[width=0.48\textwidth]{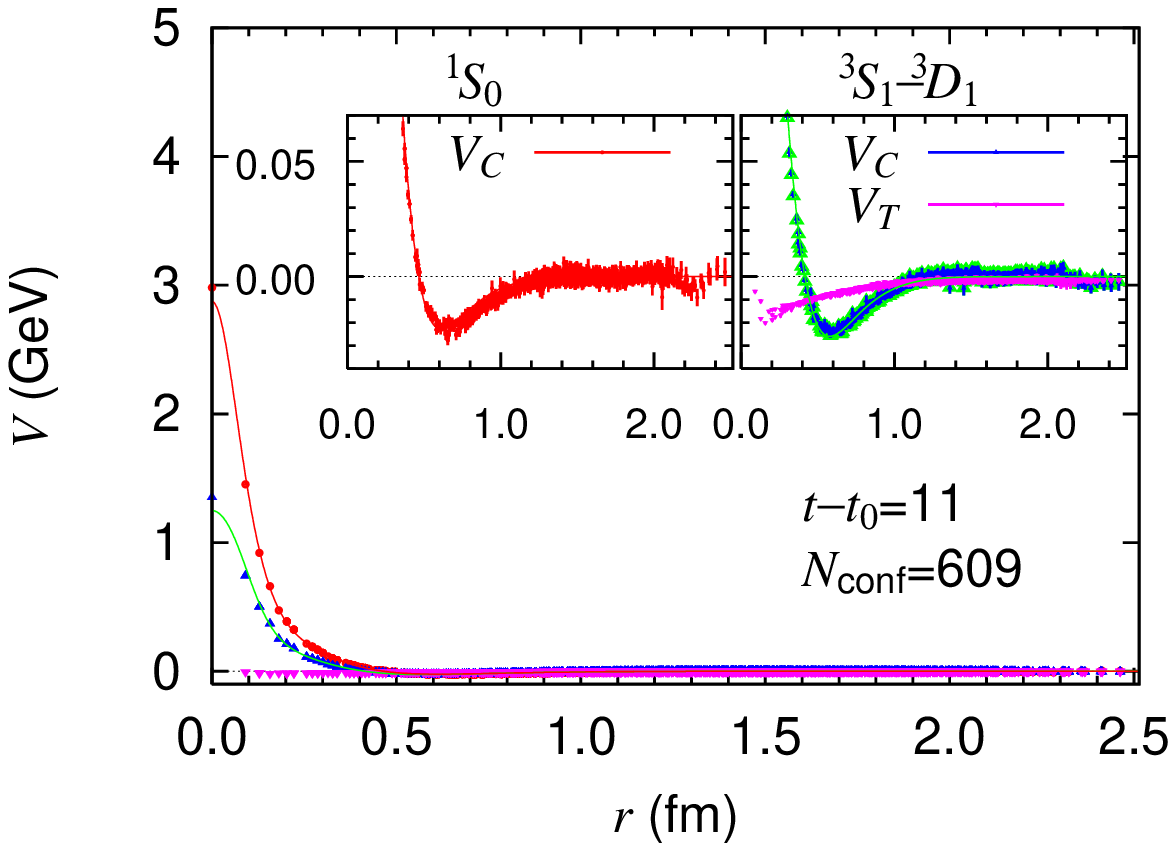}
\end{center}
\caption{$N\Xi$  potentials($I=1$)   from  quenched  QCD   (left)  and
  $N\Lambda$ potentials from 2+1 flavor QCD (right).}
\label{fig.hyperon}
\end{figure}

Hyperon potentials (hyperon-nucleon  and hyperon-hyperon) serve as the
starting  point in  studying  the hyper-nuclei  structure.  They  have
large influence on the hyperon matter generation in neutron star core.
%%%...................................................................
In spite of their importance, we  have only a limited knowledge of the
hyperon potentials,  because of the lack  of experimental information.
%% This is due  to the difficulty to perform  a direct hyperon-hyperon
%% and hyperon-nucleon scattering experiment.
%%%...................................................................
Since we do  not need any information from  the scattering experiment,
we apply  our method to  construct the hyperon potentials.   The first
attempts  have been made  to construct  $N\Xi$ ($I=0$)  potentials and
$N\Lambda$.    The  results  are   shown  in   \Fig{fig.hyperon}  (See
Refs.~\cite{nemura,nemura2} for detail.)
%%%...................................................................
These   calculations    are   being   extended    to   $N\Sigma$   and
$\Lambda\Lambda$ potentials.
%%%...................................................................
%% {\bf  Needless to  say, the  convergence of  the  derivative expansion
%%   should be examined for these results.
%% %%%...................................................................
%% However, our  method may serve  as an alternative method  to construct
%% the various hyperon potentials; @@@
%% }
It  is  important  to   examine  the  convergence  of  the  derivative
expansion.
%%%...................................................................
%% After  this  process, the  potentials  can  be  used safely  in  their
%% well-defined applicability region.
%%%...................................................................
After this process, the reliability  of the potential is guaranteed in
the well-defined applicability region.
%%%...................................................................
Such  potentials  may  be   used  as  alternatives  to  experimentally
constructed  potentials, before  experimental  information on  various
hyperon scatterings becomes fully available.

%%%...................................................................
Hyperon potentials give us another benefit.  Their flavor structure is
expected to  provide us with an  important key to  unveil the physical
origin of the repulsive core.
%%%...................................................................
To obtain  a simplified picture,  investigations in the  flavor SU(3)
idealized limit would be useful and interesting. Works are in progress
along this line \cite{inoue}.
%%%...................................................................
It would be  also interesting to compare these  numerical results with
analytical   ones   obtained  by   the   operator  product   expansion
\cite{aoki}.

\section{Summary}

We  have reported  lattice study  of nuclear  potentials based  on the
equal-time Bethe-Salpeter (BS) wave function for two nucleon system.
%%%...................................................................
We have  defined the nuclear potential by  the effective Schr\"odinger
equation demanding  that it should  generate the BS wave  functions in
wide range of energy region simultaneously.
%%%...................................................................
The  method generates  realistic  nuclear potentials,  because of  the
remarkable similarity in the  asymptotic behaviors between the BS wave
function of  QCD and the non-relativistic wave  function of scattering
state in the quantum mechanics.
%%%...................................................................
The  central and  the  tensor  potentials have  been  obtained at  the
leading order of the  derivative expansion, which show the qualitative
features of the phenomenological nuclear potentials.
%%%...................................................................
The  convergence of  the  derivative expansion  has  been examined  by
comparing  two potentials  generated at  different energies.   We have
found that the  discrepancy is small in the low  energy region $0 \alt
E_{\rm CM} \alt 45$ MeV, which indicates that the derivative expansion
works.
%%%..................................................................
For  quantitative  applications to  nuclear  physics,  2+1 flavor  QCD
should be used  to generate the nuclear potentials  in the light quark
mass region.  By using PACS-CS gauge configurations, we have attempted
to  obtain   2+1  flavor  QCD  results  of   the  nuclear  potentials.
Qualitative features  remains the same except for  the enhancements of
the  repulsive  core, the  range  of  the  attraction of  the  central
potentials, and the strength of the tensor potential.
%%%..................................................................
These nuclear potentials have been used to calculate the phase shifts,
which behave reasonably.
%%%..................................................................
Although they are reasonable in a qualitative sense, their strength is
not satisfactory at all, which  suggests the importance of the lattice
QCD calculation in the light quark mass region.
%%%..................................................................
Finally, we have applied our  method to the hyperon potentials such as
$N\Xi$ and $N\Lambda$, for which only a limited number of experimental
information is available for the moment.

It is interesting  to use our nuclear potentials  to study the nuclear
many body  problems, which  provides a way  to access nuclei  based on
QCD.
%%%...................................................................
Needless  to say,  there is  another direct  way to  access  nuclei by
lattice  QCD, i.e.,  direct calculations  of nuclear  spectrum, matrix
elements, etc \cite{yamazaki}.
%%%...................................................................
%%%...................................................................
These two approaches are considered to be complementary.
%%%...................................................................
The  former keeps  a connection  to the  conventional  nuclear theory,
while a number of formalisms and techniques have to be established.
%%%...................................................................
The  latter loses  a connection  to the  conventional  nuclear theory,
while many of the existing techniques in lattice QCD can be used.
%%%...................................................................
It is desirable  to use both of these two  approaches as the situation
demands.
%%%...................................................................
All that is certain is that the lattice QCD will provide a unique tool
to study realistic nuclei in the quite near future.

\section*{Acknowledgments}
%%%...................................................................
Quenched  QCD Monte Carlo  calculations have  been performed  with IBM
Blue Gene/L at KEK under  the ``{\em Large scale simulation program}''
at KEK (No. 09-23).
%%%...................................................................
2+1 flavor  lattice QCD Monte  Carlo calculations have been  done with
the super computer PACS-CS  and T2K under the ``{\em Interdisciplinary
Computational Science  Program}'' of Center  for Computational Science
at University of Tsukuba (No 09a-11).
%%%...................................................................
We   are  grateful   for   the  authors   and   maintainers  of   {\it
CPS++}\cite{cps}, of which a  modified version is used for measurement
done in this work.
%%%...................................................................
This  work  was  partly  supported  by Grant-in-Aid  of  the  Japanese
Ministry  of Education,  Science  and Technology,  Sports and  Culture
(Nos.
%%%...................................................................
19540261, % ishii kiban(C)
%%%...................................................................
20340047 % aoki kiban(B)
%%%...................................................................
),
%%%...................................................................
Scientific Research on Priority Areas (No. 20028013), % nemura
%%%...................................................................
Scientific  Research  on Innovative  Areas  (Nos. 20105001,  20105002,
20105003, 21105515),
%%%...................................................................
and Specially Promoted Research (No. 13002001).

\end{document}